\documentclass[journal]{IEEEtran}

\usepackage{amsfonts,amssymb,amsmath}
\usepackage{algorithm,algorithmic}
\usepackage{booktabs}
\usepackage{graphicx}  
\usepackage{cite}
\usepackage{array}
\usepackage[caption=false,font=normalsize,labelfont=sf,textfont=sf]{subfig}
\usepackage{textcomp}
\usepackage{stfloats}
\usepackage{url}
\usepackage{mathrsfs}
\usepackage{siunitx}
\usepackage{multirow} 
\usepackage[colorlinks,linkcolor=red,citecolor=blue]{hyperref}
\usepackage{pifont}
\allowdisplaybreaks[4]

\usepackage{theorem}
\newtheorem{theorem}{Theorem}

\newtheorem{definition}{Definition}
\newtheorem{assumption}{Assumption}

\newtheorem{lemma}{Lemma}

\makeatletter\def\title#1{\gdef\@title{\fontsize{22pt}{28pt}\selectfont#1}} \makeatother

\begin{document}

\title{Microservice Deployment in Space Computing Power Networks via Robust Reinforcement Learning}

\author{Zhiyong Yu,~\textit{Student Member, IEEE},
Yuning Jiang, \textit{Member, IEEE},  
Xin Liu, \textit{Member, IEEE}, \\
Yuanming~Shi, \textit{Senior Member, IEEE},
Chunxiao Jiang, \textit{Fellow, IEEE}, and 
Linling Kuang, \textit{Member, IEEE}

\thanks{Z. Yu, X. Liu, and Y. Shi are with the School of Information Science and Technology, ShanghaiTech University, Shanghai 201210, China (e-mail: {\tt \{yuzhy2023, liuxin7, shiym\}@shanghaitech.edu.cn}).}

\thanks{Y. Jiang is with the Automatic Control Laboratory, EPFL, Laussane 1015, Switzerland (e-mail: {\tt yuning.jiang@ieee.org}).}

\thanks{Chunxiao Jiang and Linling Kuang are with the Beijing National Research
Center for Information Science and Technology, Tsinghua University, Beijing,
100084, China (e-mail: {\tt \{jchx, kll\}@tsinghua.edu.cn}).}

}

\maketitle

\begin{abstract}
With the growing demand for Earth observation, it is important to provide reliable real-time remote sensing inference services to meet the low-latency requirements. The Space Computing Power Network (Space-CPN) offers a promising solution by providing onboard computing and extensive coverage capabilities for real-time inference. This paper presents a remote sensing artificial intelligence applications deployment framework designed for Low Earth Orbit satellite constellations to achieve real-time inference performance. The framework employs the microservice architecture, decomposing monolithic inference tasks into reusable, independent modules to address high latency and resource heterogeneity. This distributed approach enables optimized microservice deployment, minimizing resource utilization while meeting quality of service and functional requirements. We introduce Robust Optimization to the deployment problem to address data uncertainty. Additionally, we model the Robust Optimization problem as a Partially Observable Markov Decision Process and propose a robust reinforcement learning algorithm to handle the semi-infinite Quality of Service constraints. Our approach yields sub-optimal solutions that minimize accuracy loss while maintaining acceptable computational costs. Simulation results demonstrate the effectiveness of our framework.

\end{abstract}


\section{Introduction}
\label{Section I}

Remote sensing (RS) satellite system, with multiple downstream tasks such as environmental and disaster monitoring, plays a significant part in Earth observation missions through global coverage, all-weather, and full-spectrum detection capacities. However, the vast volume of high-resolution raw data captured by the RS satellites raises a challenge: directly downloading raw data to ground stations for processing would result in critical latency issues \cite{10605604}. The Space Computing Power Network (Space-CPN) is a promising solution to address this problem by integrating communication and inference capabilities into satellite constellation networks \cite{zhu2024hierarchical}. In particular, the Space-CPN is a multilayer satellite-based distributed computing platform, enabling RS tasks to perform similarly to ground-based data centers. Recent studies \cite{bui2023board}, \cite{leyva2023satellite}, \cite{koubaa2023aero} focus on the Low Earth Orbit (LEO) onboard inference within the Space-CPN, and the objective is to minimize the total service latency and to achieve the better inference performance.

\begin{figure*}[t]
\centering
\includegraphics[width=\linewidth]{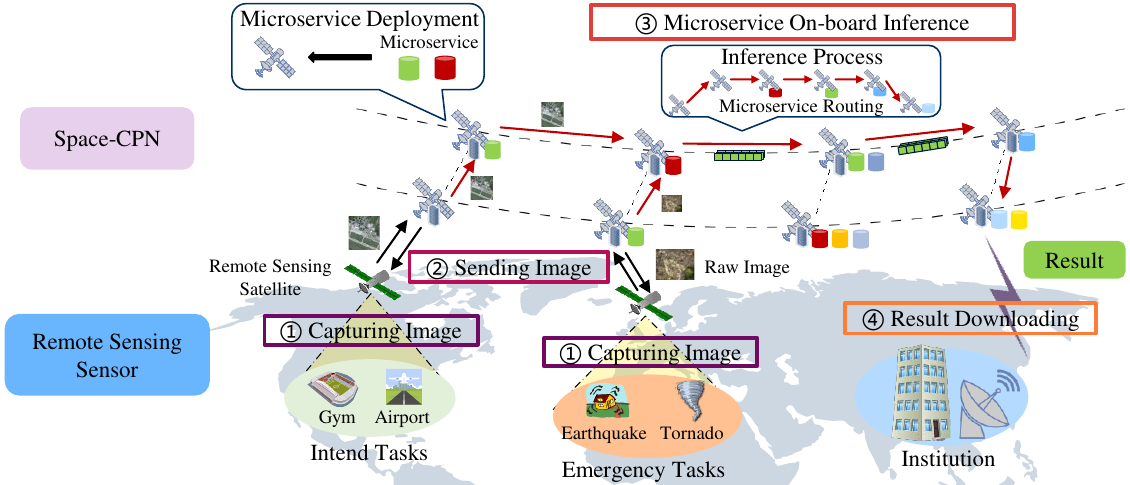}
\caption{On-board inference process of remote sensing.}
\label{Fig. 1}
\end{figure*}

Although the Space-CPN enables onboard inference capabilities for the RS applications, monolithic inference still faces multiple challenges. On the one hand, RS tasks such as land use category classification and disaster monitoring require high computation ability that a single LEO satellite cannot afford \cite{zhang2024earthgpt}. On the other hand, these downstream tasks often involve redundant module deployment and computation, which results in inefficiencies \cite{wang2020mpcsm}. We adopt the microservice architecture in our on-board inference framework to address these challenges, which divides the monolithic application into multiple low-coupled modules (i.e., microservices) \cite{jang2021microservice}. These microservices are deployed on suitable satellite nodes to perform the monolithic application's function. This architecture takes advantage of the portability, scalability, and resilience in software engineering. As illustrated in Fig. \ref{Fig. 1}, the Space-CPN's satellite will receive image data from the RS satellite and start the on-board inference to send the result to the ground station. Thus, an efficient microservice deployment strategy is necessary for the inference to proceed smoothly.



However, the challenge arises from the inherent heterogeneity of satellites because each satellite has different computing and communication capacities \cite{zhang2022aerial}. In particular, several satellites in Space-CPN serve as the communication relay with few computing and storage resources, and several satellites with high-performance hardware enable high computation demand tasks. Therefore, the heterogeneity property necessitates the deployment algorithm to select the most suitable satellite for each microservice to maximize resource utilization and ensure efficient inference (i.e., reduce latency punishment) \cite{shi2023task}, \cite{letaief2019roadmap}. Consequently, the onboard microservice deployment problem becomes NP-hard with numerous local optima due to the LEO network's mesh topology, non-convex objectives, and integer constraints \cite{wang2020mpcsm}, \cite{mazyavkina2021reinforcement}. 
The articles \cite{9615028}, \cite{9162056}, \cite{10162207} use reinforcement learning (RL) based algorithms to solve the microservice deployment problem on the base station with powerful computing ability. \cite{9740415} formulated a fractional polynomial problem caused by multiple instances of a single microservice and proposed a greedy-based heuristic algorithm to solve it. \cite{10013701} considered the interference between microservices competing for the same resources and proposed a low-complexity heuristic algorithm with parallel deployment ability to minimize the use of the servers. However, these works did not consider LEO satellite constellation properties, i.e., low computing power and few resources. \cite{liu2019e3}, \cite{sami2020vehicular}, \cite{9507367} try deploying microservice on edge devices such as Nvidia Jetson to meet the load balance, low latency, and security constraints. Although these works take advantage of low-power platforms, they cannot meet the LEO satellite constellation's topology properties and may receive critical resource waste and service latency punishment. \cite{yan2019satec}, \cite{su2023attention} proposed the satellite onboard deployment framework and algorithm to meet the constraints of LEO satellite constellation, but deterministic modeling cannot meet the uncertainty of user requests from different regions. Thus, an efficient, robust microservice deployment algorithm is necessary to reduce total resource consumption with Quality of Service (QoS) constraints and meet resource requirements. 

The main challenge for the deterministic model when inference in the natural environment is that satellites need to complete the pre-assigned tasks established at launch and be ready to process emergency tasks such as earthquake and forest fire monitoring. These emergency tasks often occur unexpectedly with uncertainties in time and frequency \cite{zhai2015robust}. To tackle this, we employ the Robust Optimization (RO) approach to capture the uncertainty in the data amount \cite{balouka2021robust}. However, the dependency relationships between microservices complicate the deployment problem by increasing the coupling between optimization variables. Moreover, the semi-infinite QoS constraints complicate the optimization problem. As a result, traditional solvers like \texttt{Complex} or \texttt{Gurobi} cannot solve this problem directly. To address this, we reformulate the deterministic problem as a Partially Observable Markov Decision Process (POMDP) and solve it with reinforcement learning (RL) because of its MDP characteristics, which allows us to make decisions with partial knowledge of the optimization problem. \cite{10.5555/3305890.3305972}, \cite{10.5555/3495724.3497489}, \cite{rigter2022ramborl} using adversaries to perturb the state the protagonist agent observes to receive a robust strategy. \cite{wang2021online}, \cite{wang2023policy}, \cite{pmlr-v162-wang22at} derived the agent's updating equation with R-contamination uncertainty set on the transition kernel. \cite{he2023robust} introduced uncertainty modeling into multi-agent reinforcement learning and analyzed the two agents’ adversarial equilibrium conditions. Therefore, it is necessary to propose an efficient and robust deployment algorithm based on robust RL to minimize resource consumption and enhance the system's inference ability. 

\begin{figure*}[t]
\centering
\includegraphics[width=\linewidth]{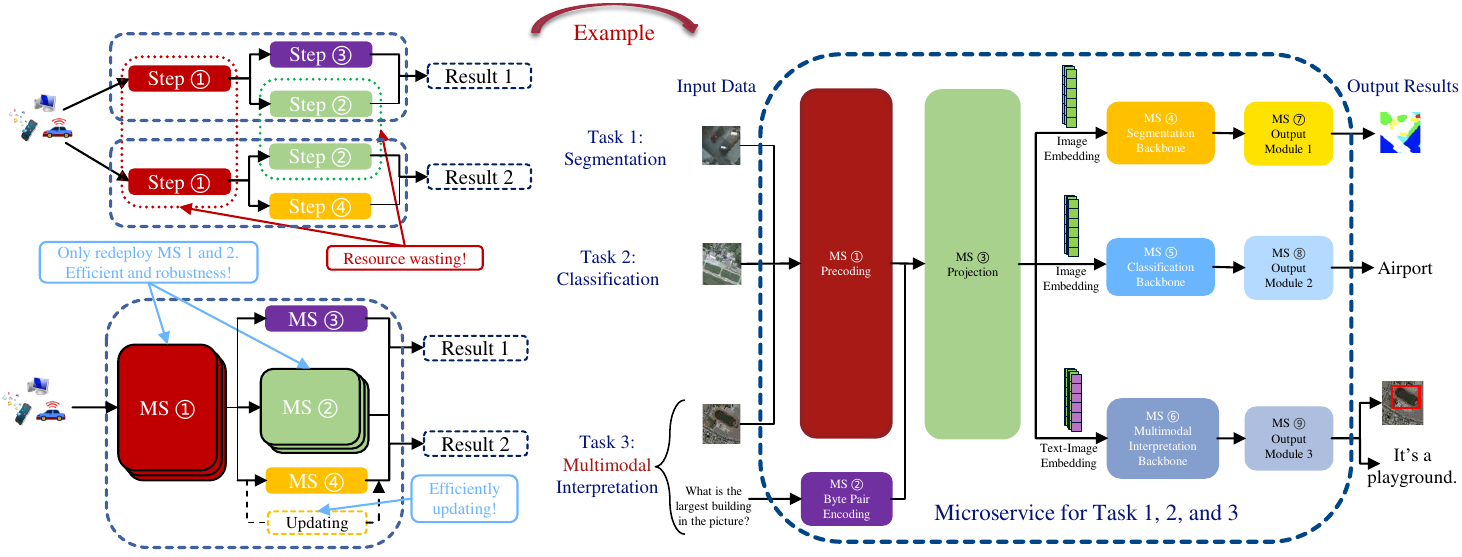}
\caption{Remote sensing onboard satellite edge AI inference with microservice architecture.}
\label{microservices framework}
\end{figure*}

\subsection{Contributions}
In this paper, we proposed a robust reinforcement learning-based inference framework for resource consumption minimization in satellite onboard microservice deployment. The major contributions are summarized as follows:
\begin{itemize}
\item A robust on-board microservice deployment focusing on the user request uncertainty is proposed to minimize the LEO satellite constellation resource consumption and the QoS punishment. It uses a box uncertainty set on the data amount from each region to control the robust level (i.e., the number of microservices deployed), thereby making a trade-off between minimizing the resource consumption and avoiding the QoS punishment. Moreover, we incorporate explicit QoS constraints for each request to ensure that all inference tasks are processed within low latency.

\item We first decompose the original problem into two parts, based on the redeployment costs associated with each microservice. We use the traditional reinforcement learning approach in the first stage to deploy the core microservices on the LEO satellite as the data center. In the second stage, we reformulate the optimization problem into a Partly Observable Markov Decision Process (POMDP) and solve it by using robust reinforcement learning. Due to the unexpected perturbation of POMDP, we use robust adversarial reinforcement learning to solve this robust microservice deployment problem. We analyze the equilibrium of this adversarial game and prove the existence of Nash equilibrium in this game. Simulation results will demonstrate the effectiveness of our proposed robust microservice deployment framework in minimizing resource consumption and avoiding QoS punishment.

\end{itemize}

\subsection{Organization}
The remainder of this paper is organized as follows. Section \ref{Section II} proposes the microservice model, LEO satellite model, and deterministic deployment problem of LEO satellite constellation and microservice. In Section \ref{Section III}, the robust optimization problem with semi-infinite QoS constraints is proposed to formulate the microservice deployment with data amount uncertainty. In Section \ref{Section IV}, we transfer the original problem into two MDPs and propose two microservice deployment algorithms to solve this problem. Simulations and discussions are given in Section \ref{Section V}. Finally, conclusions are presented in Section \ref{Section VI}.

\subsection{Preliminaries and Notations}
\label{PAN}
In this paper, we use bold uppercase letters to represent matrices, while bold lowercase letters represent vectors. 
For a given set $\mathbb X$, we use notation $\lvert\mathbb X\rvert$ to denote its cardinality. $\mathcal{I}(\cdot)$ is the indicator function which equals to 1 when equation in $(\cdot)$ is true and equal to 0 for others, $\text{vec}(\cdot)$ denotes the vectorization operation. Besides, the most commonly used notations are listed as Table \ref{table1}. For a given two-player zero-sum game, we denote it by $\mathbb{G}=\{\{1,2\},\{\mathbb{S}_{1},\mathbb{S}_{2}\},\{\mathbb{Y}^{t},\mathbb{Z}^{t}\},\mathbf{C}\}$, where $\mathbb{S}_{1}$ and $\mathbb{S}_{2}$ denote two players' strategy set respectively, $\mathbb{Y}^{t}$ and $\mathbb{Z}^{t}$ denote probability set of each strategy in $\mathbb{S}_{1}$ and $\mathbb{S}_{2}$ respectively, $\mathbf{C}\in\mathbb{R}^{\lvert\mathbb{Y}^{t}\rvert\times\lvert\mathbb{Z}^{t}\rvert}$ denotes the reward matrix of two players. Let 
\begin{align}
    v_{1}=\max_{\mathbf{y}\in \mathbb{Y}^{t}}\min_{\mathbf{z}\in \mathbb{Z}^{t}}\sum_{i=1}^{|\mathbb{Y}^{t}|}\sum_{j=1}^{|\mathbb{Z}^{t}|}c_{ij}y_{i}z_{j},\\
    v_{2}=\min_{\mathbf{z}\in \mathbb{Z}^{t}}\max_{\mathbf{y}\in \mathbb{Y}^{t}}\sum_{i=1}^{|\mathbb{Y}^{t}|}\sum_{j=1}^{|\mathbb{Z}^{t}|}c_{ij}y_{i}z_{j},
\end{align}
we can have the following definition for Minmax Equilibrium.
\begin{definition}[Minimax Equilibrium]
A minimax equilibrium existing in the two-player zero-sum game is equivalent to $v_{1}=v_{2}$.
\end{definition}

\section{System~Model}
\label{Section II}
In our onboard inference system, it is necessary to consider the topology of the LEO satellite constellation and the inference scheme. To this end, Section \ref{MM} and \ref{LEOM} present the model of the microservice and the LEO constellation. The latency model, cost model, and constraint of microservices are introduced in Section \ref{LM}, \ref{TSMCM}, and \ref{Constraint}, respectively. Then we propose the deterministic optimization problem in Section \ref{DPF}.

\subsection{Microservice Model}
\label{MM}
Microservice architecture (MSA) is a programming paradigm for decomposing applications into a collection of light, independent microservices (MS). Each MS runs in its own process and communicates with other modules.
\begin{table}
\footnotesize
\renewcommand{\arraystretch}{1.3}
\label{table1}
\caption{The notation used in this article}
\begin{tabular}{|c|c|}
\hline
\textbf{Notation} & \textbf{Description} \\
\hline
$O,L$ & Amount of orbital planes and satellites in each plane \\
\hline
$D_{u,v}$ & Distance between satellite $u$ and $v$ \\
\hline
$G_{D},G_{M}$ & Graph of LEO constellation and microservices \\
\hline
$\mathbb{M}_{l}, \mathbb{M}_{c}, \mathbb{M}$ & Edge, core, and whole satellite set \\
\hline
$\mathbb{E}_{1}$ & Set of data flow between each microservice \\
\hline
$\mathbb{D}$ & Set of available satellites \\
\hline
$\mathbb{E}_{2}$ & Set of ISLs between two satellites \\
\hline
$R$ & The number of resource types \\
\hline
$r_{d}(i),r_{m}(i)$ & Resource variable of satellite $d$ and microservice $m$ \\
\hline
$\mathbb{A}$ & Set of microservices' computing demand \\
\hline
$\mathbb{B}$ & Set of result data amount generated by microservices \\
\hline
$p^{f}_{m},p^{k}_{m},p^{ct}_{m}$ & Consumption of deployment, maintaining, parallel \\
\hline
$\tau^{tr}_{u,i,j}$ & Transmission latency of $u$ between satellite $i,j$ \\
\hline
$\tau^{pp}_{u,i,j}$ & Propagation latency of $u$ between satellite $i,j$ \\
\hline
$\tau^{pc}_{u,i}$ & Processing latency of $u$ on satellite $i$ \\
\hline
$L$ & Speed of light \\
\hline
$W_{i,j}$ & Data transmission rate between satellite $i,j$\\
\hline
$f_{i}$ & Computing speed of satellite $i$ \\
\hline
$l$ & Total latency of all microservices \\
\hline
$\mathbb{T}$ & Total service time slot set \\
\hline
$\mathbf{X},\mathbf{Y}$ & Deployment scheme matrix of two microservices sets \\
\hline
$\mathcal{C}_{1}(\mathbf{X})$ & Total core microservices deployment cost \\
\hline
$\mathcal{C}_{2}(\mathbf{Y})$ & Total light microservices deployment cost \\
\hline
$\mathcal{C}_{3}(\mathbf{Y})$ & Total light microservices maintaining cost \\
\hline
$\mathcal{C}_{4}(\mathbf{Y})$ & Total light microservices parallel cost \\
\hline
$\mathbf{Z}$ & Request matrix of all time slots \\
\hline
$\Omega(\mathbf{Z}^{t})$ & Uncertainty set on $\mathbf{Z}^{t}$ \\
\hline
$p_{u,d}$ & Microservice $u$ is deployed on satellite $d$ or not \\
\hline
$s^{p}_{i},s^{d}_{i}$ & protagonist and adversary state\\
\hline
$a^{p}_{i},a^{d}_{j}$ & protagonist and adversary action\\
\hline
$\Phi$ & Width of uncertainty set \\
\hline
$r^{p}_{i},r^{d}_{j}$ & protagonist and adversary reward\\
\hline
\end{tabular}
\end{table}
In a monolithic architecture, we must redeploy several components multiple times for all tasks, resulting in inefficient resource consumption. Additionally, the high coupling of monolithic systems makes it difficult to update individual components, as any update requires redeploying the entire application. To address this, the remote sensing microservice inference system we considered in this paper is shown in Fig. \ref{microservices framework}. In particular, there exist 3 tasks: segmentation, classification, and multimodal interpretation and 9 microservices: \ding{182} precoding, \ding{183} byte pair encoding, \ding{184} projection, \ding{185} segmentation backbone, \ding{186} classification backbone, \ding{187} multimodal interpretation backbone, and \ding{188}$\sim$\ding{190} output module 1$\sim$3. Thus, the inference tasks are shown as: \normalsize{\textcircled{\scriptsize{1}}} segmentation (\ding{182}$\rightarrow$\ding{184}$\rightarrow$\ding{185}$\rightarrow$\ding{188}), \normalsize{\textcircled{\scriptsize{2}}} classification (\ding{182}$\rightarrow$\ding{184}$\rightarrow$\ding{186}$\rightarrow$\ding{189}), and \normalsize{\textcircled{\scriptsize{3}}} multimodal interpretation (\ding{182}, \ding{183}$\rightarrow$\ding{184}$\rightarrow$\ding{187}$\rightarrow$\ding{190}) where $\rightarrow$ denotes the data flow between the microservices. Due to the low coupling, we can update the core components such as \ding{185}$\sim$\ding{187} with no system shutdown and reduce the service waiting latency. Due to the tasks \normalsize{\textcircled{\scriptsize{1}}}, \normalsize{\textcircled{\scriptsize{2}}}, and \normalsize{\textcircled{\scriptsize{3}}} are using the same microservices \ding{182} and \ding{184}, the MSA system reuses these microservices to reduce the resource consumption.

To leverage the computational power of Space-CPN, we deploy the remote sensing inference system using MSA within the LEO constellation. As illustrated in Fig. \ref{Fig. 1}, the on-board inference pipeline operates as follows: Microservices are distributed across the constellation within Space-CPN. During inference, satellites capture high-resolution images for routine tasks (e.g., monitoring gyms and airports) and emergency tasks (e.g., detecting earthquakes and tornadoes). These images are transmitted to the constellation, where microservices process them and transfer data via a predefined routing mechanism. The results are then sent to ground stations for delivery to repositories like government agencies or research institutions. Deploying all microservices on a single satellite is infeasible due to the heterogeneous computational and resource capabilities across satellites. Furthermore, deploying every microservice on all satellites is impractical, as the large data files associated with multiple inference backbones exceed storage capacity.

Let $G_{M}=(\mathbb{M},\mathbb{E}_{1})$ denotes the directed acyclic graph of remote sensing application where $\mathbb{M}$ is the set of microservices required by the application and $\mathbb{E}_{1}$ represent the data flow between each microservice. Due to the inference model such as Llama \cite{touvron2023llama} needing a significantly large parameter file (over 4 GB), the redeployment latency is unacceptable because of the rate of ISL \cite{nguyen2021design}. Let $\mathbb{M}_{l}$ (i.e., precoding, byte pair encoding, and projection) denote the set of the light microservices, and $\mathbb{M}_{c}$ (i.e., segmentation, classification, multimodal interpretation backbone, and output model 1$\sim$3) denote the set of the core microservices. Thus, the set of all microservices in this application can be depicted as $\mathbb{M}=\mathbb{M}_{l}\cup \mathbb{M}_{c}$. For arbitrary microservice $m \in \mathbb{M}$, we use $r_{m}(i)\ge 0,\;i=0,1,...,R$ to denote the microservice's requirement of the corresponding resource $i$ where $R$ denote the number of resource types. We use $a_{m}>0$ to denote the computing data amount (bits) of microservice $m$ and $\mathbb{A}=\{a_{1},a_{2},...,a_{M}\}$ with $M=|\mathbb{M}|$. And we use $b_{m}$ to denote the data amount of microservice $m$'s result (bits) and $\mathbb{B}=\{b_{1},b_{2},...,b_{M}\}$ to denote the set of result's data amount of all microservices.

\subsection{LEO Constellation Model}
\label{LEOM}
A LEO satellite constellation functions similarly to a ground-based data center, providing real-time support for onboard remote sensing applications. A typical LEO constellation with Walker Star construction, such as Iridium, consists of $O$ orbital planes, and each orbital contains $L$ satellites. It can be depicted as a grid graph $G_{D}=(\mathbb{D},\mathbb{E}_{2})$, where $\mathbb{D}$ is the set of available satellites (i.e., $D=\lvert \mathbb{D} \rvert = O\times L$), and $\mathbb{E}_{2}$ is the set of ISLs between two satellites.  Each satellite in the constellation has 4 inter-satellite links (ISLs), connecting two satellites within the intra-plane and two satellites in the inter-plane. Due to the Earth's obstruction and limited ISLs, one satellite can only transfer data by multi-hop with satellites without ISLs. Distance $D_{u,v}$ (kilometers) between satellite $u$ and $v$ is determined by the latitudes, longitudes, and altitudes of two satellites. $D_{u,v}$ can be depict as follows \cite{deng2022distance}:
\begin{subequations}
\label{eq::distance}
\begin{align}
D_{u,v}&=\sqrt{2r^{2}[1-\cos\phi_{u}\cos\phi_{v}\cos(\lambda_{u}-\lambda_{v})-\sin\phi_{u}\sin\phi_{v}]}, \label{duv}\\
\phi&=\arcsin(\sin i\sin \mu), \label{xu}\\
\lambda&=\Omega +\arctan(\cos i \tan \mu) -\omega_{2}t, \label{yu}\\
\mu&=\omega_{1}t+\gamma, \label{zu}
\end{align}
\end{subequations}
where $r$ denotes the sum of the Earth's radius and the satellite's orbit altitude (in kilometers), $i$ is the inclination of the LEO satellite orbit, and $\Omega$ is the position of the ascending node, measured in degrees. Additionally, $\omega_{1}$ represents the satellite's angular velocity as it orbits the Earth, with units of degrees per millisecond. $\gamma$ is the initial phase angle of the satellite. $\omega_{2}$ is the angular velocity of the earth's rotation in degrees/milliseconds. $t$ is the current time. Let $\mathbb{R}$ be the set of available satellite resources (e.g., CPU core, GPU core, memory, power). For each satellite $d\in\mathbb{D}$, we use $r_{d}(i)\ge 0,\;i=0,1,...,R$ to denote the satellite's providing capacity of the corresponding resource $i$. We use $c_{d}>0$ to denote the computing ability (bits/ms) of satellite $d$ and $\mathbb{C}=\{f_{1},f_{2},...,f_{D}\}$ denote the set of satellites' computing ability (bits/ms) with $D=|\mathbb{D}|$.

\subsection{Latency Model}
\label{LM}
Due to the dependency of microservices, the total latency (milliseconds) of microservice $v$ at time slot $t$, which includes processing, propagation, and transmission latency, is influenced by the performance of its preceding microservice $u$ \cite{cao2023computing}. Therefore, the latency of microservice $v$ is defined as
\begin{align}\label{ttt}
\tau_{v}(u,i,j)=\tau^{tr}_{u,i,j}+\tau^{pp}_{u,i,j}+\tau^{pc}_{v,j},
\end{align}
where $u,v$ deployed on satellite $i,j$ respectively. $\tau^{tr}_{u,i,j}$ and $\tau^{pp}_{u,i,j}$ denote the transmission latency and the propagation latency of microservice $u$ from satellite $i$ to satellite $j$ respectively. $\tau^{pc}_{v,j}$ denotes the processing latency of microservice $v$ deploy on satellite $j$.

Specifically, the propagation latency occurring from microservice $u$ to $v$ is associated with the spatial distance between the satellites of $u$ and $v$, as well as the speed of light. Thus, the propagation latency can be defined by
\begin{align}\label{tpp}
\tau^{pp}_{u,i,j}=D_{i,j}/L,
\end{align}
where $\tau^{pp}_{u,i,j}$ denotes the propagation latency of microservice $u$ from satellite $i$ to satellite $j$, $D_{i,j}$ represents the distance (kilometers) between $i$ and $j$ at the time the message is sent, $L$ denotes the speed of light (kilometers/ms). The transmission latency is related to the distance between two satellites and the total data generated by microservice $u$. The transmission latency from microservice $u$ to $v$ can be depicted by
\begin{align}\label{ttr}
\tau^{tr}_{u,i,j}=b_{i}/W_{i,j},
\end{align}
where $W_{i,j}$ denotes the data transmission rate (bits/ms) between satellite $i$ and $j$. Besides, if microservice $u$ is the first microservice of an application, the propagation latency $\tau^{pp}_{u,i,j}=0$. The processing latency is related to the total amount of computation generated and the satellites' computational ability (CPU frequency, GPU float ability, and so on). So, the processing latency of microservice $v$ deployed on the satellite $i$ can be defined as
\begin{align}\label{tpc}
\tau^{pc}_{v}=a_{v}/f_{i},
\end{align}
where $f_{i}\in \mathbb{C}$ is the computation ability (bits/ms) of satellite $i$. Therefore, the total latency of the application is defined as
\begin{align}\label{l}
l=\sum_{m\in \mathbb{M}}\tau_{m}.
\end{align}
Besides, if microservice $m$ has a dependency relation with more than one microservice, $\tau_{m}$ will choose the arrival time of the last received microservice $u$'s data to compute the $\tau^{tr}$ and $\tau^{pp}$.

\subsection{Microservice Cost Model}
\label{TSMCM}
We denote the microservice deployment scheme of time slot $t$ as two matrixes $\mathbf{X}^{t}\in \mathbb{N}^{M_{c}\times D},\mathbf{Y}^{t}\in \mathbb{N}^{M_{l}\times D}$, where each entry is the deployment count of microservice $m$ on satellite $s$, $M_{l}=\lvert\mathbb{M}_{l}\rvert$, $M_{c}=\lvert\mathbb{M}_{c}\rvert$. Let $\mathbf{X}=[\mathbf{X}^{0},\mathbf{X}^{1},...,\mathbf{X}^{T}], \mathbf{Y}=[\mathbf{Y}^{0},\mathbf{Y}^{1},...,\mathbf{Y}^{T}]$ denote the deployment scheme of core microservices and light microservices, respectively, where $T=\lvert\mathbb{T} \rvert$, $\mathbb{T}$ denotes the total service time set. 
The total cost (i.e. money) of the whole inference system during $T$ time slots can be divided into four partitions: money cost of core microservices $\mathcal{C}_{1}(\mathbf{X})$, deployment money cost of light microservices $\mathcal{C}_{2}(\mathbf{Y})$, maintenance money cost of light microservices $\mathcal{C}_{3}(\mathbf{Y})$, and parallel computing money cost of light microservices $\mathcal{C}_{4}(\mathbf{Y})$ \cite{nguyen2021two}. The money cost of core microservices is influenced by the deployment counts, which are shown as follows:
\begin{align}\label{o1}
\mathcal{C}_{1}(\mathbf{X}) = \sum_{m\in \mathbb{M}_{c}}\sum_{s\in {\mathbb{D}}}\{p_{m}^{d}x_{m,s}^{0} + \sum_{t\in T}p_{m}^{k}x_{m,s}^{0}\},
\end{align}
where $p^{d}_{m},~p^{k}_{m}$ denotes the deployment price and a time slot's maintenance price of microservice $m$ on satellite $s$ at time slot $t$, respectively. The deployment money cost of light microservices for a time slot can be depicted as follows:
\begin{align}\label{o2}
\mathcal{C}_{2}(\mathbf{Y}) = \sum_{m\in \mathbb{M}_{l}}\sum_{s\in {\mathbb{D}}}\sum_{t\in \mathbb{T},t\neq 0}p^{d}_{m}\max(0,y^{t}_{m,s}-y^{t-1}_{m,s}),
\end{align}
where $p^{k}_{m}$ denotes the deployment price of microservice $m$ on satellite $s$ at time slot $t$. The maintenance money cost of light microservices for a time slot is shown as follows:
\begin{align}\label{o3}
\mathcal{C}_{3}(\mathbf{Y}) = \sum_{m\in \mathbb{M}_{l}}\sum_{s\in {\mathbb{D}}}\sum_{t\in \mathbb{T}}p^{k}_{m}y^{t}_{m,s},
\end{align}
where $p^{k}_{m}$ denotes the deployment price of microservice $m$ on satellite $s$ at time slot $t$. 
The parallel computing money cost of light microservices for a time slot is:
\begin{align}\label{o4}
\mathcal{C}_{4}(\mathbf{Y}) = \sum_{m\in \mathbb{M}_{l}}\sum_{s\in {\mathbb{D}}}\sum_{t\in \mathbb{T}}p^{a}_{m}y^{t}_{m,s},
\end{align}
where $p^{a}_{m}$ denotes the parallel computing price of microservice $m$ on satellite $s$ at time slot $t$.

\subsection{Deterministic Constraint}
\label{Constraint}
To meet the resource and QoS limitation, the deployment scheme will deal with the following constraints.
\subsubsection{Function Completeness Constraint}
\label{FCC}
To simulate the function of a monolithic service, each microservice should be deployed at least one time. In particular, for $x^{0}_{m,s}$, the completeness constraint and the integer constraint is:
\begin{align}
    \label{x1}
    \sum_{s\in \mathbb{D}}x^{0}_{m,s}\ge 1,\;x^{0}_{m,s}\in \mathbb{N}
\end{align}
for all $m\in \mathbb{M}_{c}, s\in \mathbb{D}$. For $y^{t}_{m,s}$, it was shown as:
\begin{align}
    \label{y1}
    \sum_{s\in \mathbb{D}}y^{t}_{m,s}\ge 1,\;y^{t}_{m,s}\in \mathbb{N}
\end{align}
for all $m\in \mathbb{M}_{l},s\in \mathbb{D},t\in\mathbb{T}$.

\subsubsection{Parallel Computing Constraint}
\label{PCC}
We use the relationship between area requests and parallel accessing ability $k_{m}$ of each microservice to model the deterministic parallel constraint. We use $\mathbf Z^{t} \in \mathbb{R}^{M\times D}$ to denote the request in time slot $t$. Each microservice can only process at most $k_{m}$ requests during a time slot, i.e., each task request $z_{m,s}^{t}\in \mathbf{Z}^{t}$ from all areas needs to find a microservice that still can process one more task request (i.e. one more raw image), which can be formulated as
\begin{align}
\label{yk}
\sum_{s\in \mathbb{D}} y_{m,s}^{t}k_{m}\ge \sum_{s\in \mathbb{D}}z_{m,s}^{t}
\end{align}
for all $t\in \mathbb{T},m\in \mathbb{M}_{l}$.

\subsubsection{QoS Constraint}
To ensure each request can be responded to efficiently, we defined QoS (Quality of Service) in our system as the total latency of each task. We use $z\in \mathbb{Z}^{t}$ to denote the index of request at time slot $t$ with 
$$\mathbb{Z}^{t}=\left\{z\in \mathbb{N}\mid 0\le z\le \sum_{s}^{\mathbb{D}}z_{0,s}^{t}\right\}.$$ To track the inference routing of each task, we use $\mathbf{R}^{t}\in \mathbb{N}^{Z \times M}$ to denote the place of microservices used by each task where $Z^{t}=\lvert \mathbf{Z}^{t}\rvert$, elements in $\mathbf{R}^{t}$ denotes the microservice's deployment place. In our system, each task will try to use the microservices routing path with the shortest distance if processing ability exists. In particular, we define the chosen $\mathbf{R}^{t*}$ by
$\mathbf{R}^{t*}:=$
\begin{align}   
\label{Rstar}
&\mathop{\arg\min}\limits_{\mathbf{R}^{t}\in \mathbb{N}^{Z^{t} \times M_{l}}}\Bigg\{ \sum_{z=0}^{Z^{t}-1}\sum_{p=1}^{M-1}\Vert\mathcal{F}(r^{t}_{z,p-1})-\mathcal{F}(r^{t}_{z,p})\Vert_2\;+ \\\notag
&\qquad \alpha_{p}\cdot \max\Big(0,\sum_{m=0}^{M_{l}-1}\sum_{d=0}^{D-1}\{\sum_{z=0}^{Z^{t}-1}\mathcal{I}(r^{t}_{z,m}=d)-y^{t}_{m,d}\}\Big)\Bigg\},
\end{align}
where $r^{t}_{z,m}\in \mathbf{R}^{t}$ denotes the microservice $m$'s deployment place used by task $z$ at time slot $t$, $\alpha_{p}$ is a large factor which denotes the overload punishment, and $(c_{1},c_{2})=\mathcal{F}(r_{1})$ is the satellite's coordinate in the mesh topology defined as $c_{1}=r_{1}/L,\;c_{2}=r_{1}\bmod L$. With the microservice routing matrix $\mathbf{R}^{t*}$ and equations~\eqref{eq::distance}, \eqref{ttt}, and~\eqref{l}, we can formulate the QoS constraint for each task as:
\begin{align}
\label{QoSC}
l_{z}=\sum_{m=1}^{M_{l}-1}\tau_{v}(m-1,r^{t}_{z,m-1},r^{t}_{z,m})\le QoS
\end{align}
for all $z\in \mathbb{Z}^{t}$, $r^{t}_{z,m} \in \mathbf{R}^{t*}$, where $QoS$ is a scalar denotes the maximum service time, $l_{z}$ denotes service time of request $z$ according to equation \eqref{l}.

\subsubsection{Core Microservices Resource Constraint}
\label{CMRC}
The total resource consumption of the core microservices deployed on one satellite cannot exceed the resource capacity of this satellite:
\begin{align}
\label{CMRCal}
    \sum_{m\in \mathbb{M}_{c}}r_{m}(j)x^{0}_{m,s}\le r_{s}(j)
\end{align}
for all $j\in \mathbb{R}, s\in \mathbb{D}$.

\subsubsection{Light Microservices Resource Constraint}
\label{EMRC}
Due to the core microservices, the light microservices are restricted by the satellite's resource capacity and the core microservices' resource consumption:
\begin{equation}
\label{EMRCal}
    \begin{aligned}
        \sum_{m\in \mathbb{M}_{l}}\left\{r_{m}(j)x^{0}_{m,s}+r_{m}(j)y^{t}_{m,s}\right\}\le r^{t}_{s}(j)
    \end{aligned}
\end{equation}
for all $j\in \mathbb{R},t\in \mathbb{T},s\in \mathbb{D}$.

\begin{figure}[t]
\centering
\includegraphics[width=\linewidth]{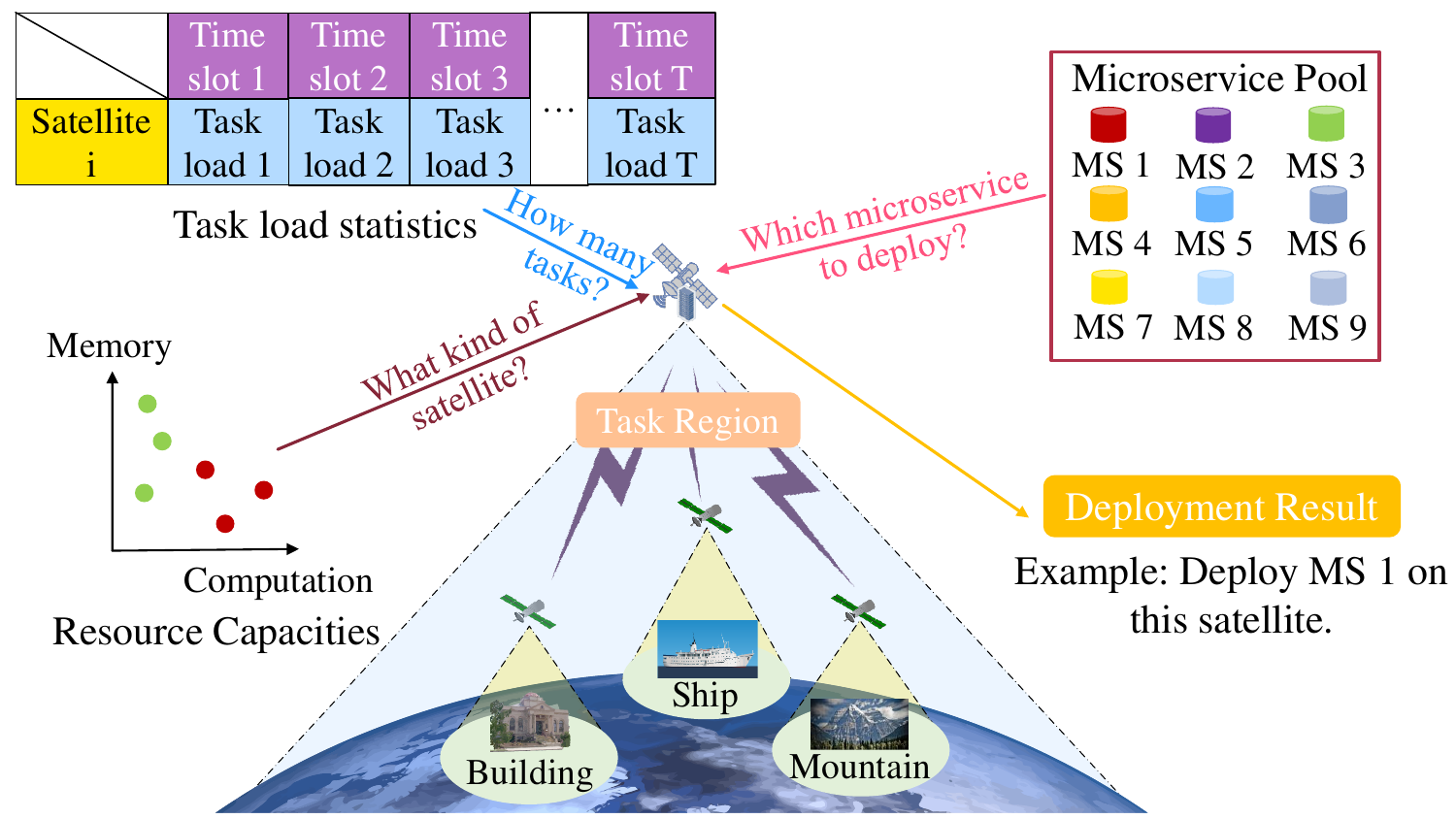}
\caption{Microservice deployment problem specification.}
\label{eomdp}
\vspace{-0.3cm}
\end{figure}

\subsection{Microservice Deployment Problem Formulation}
\label{DPF}
Microservice deployment in Space-CPN is to allocate microservices in the finite time slots (i.e., $T$ time slots) to appropriate satellites based on task load statistics, server latency, and resource (i.e., CPU, GPU, memory) constraints, which is shown in Fig. \ref{eomdp}. In particular, the deployment system will use the statistics: task load $\mathbf{Z}^{t}$ and satellite's resource $r_{m}(i)$ to decide which microservice will deploy on the current satellite. The objective is to minimize the overall money cost of all microservices in any time slot $t$ with no QoS violation. By the cost model and the constraints proposed in Section \ref{TSMCM} and Section \ref{Constraint}, we formulate the deterministic optimization problem as follows:
\begin{subequations}
\begin{align}
\label{deter}
    \min_{\mathbf{X},\mathbf{Y}}\;\;&\mathcal{C}_{1}(\mathbf{X})+ \mathcal{C}_{2}(\mathbf{Y})+ \mathcal{C}_{3}(\mathbf{Y})+ \mathcal{C}_{4}(\mathbf{Y})\\
    \text{subject to}\;\;&\text{constraints}~\eqref{x1}-\eqref{EMRCal},
\end{align}
\end{subequations}
where $\mathbf{X}\in \mathbb{R}^{M_{c}\times D}, \mathbf{Y}\in \mathbb{R}^{T\times M_{l}\times D}$ denote the deployment scheme of two kinds of microservice. 

\section{Robust~Microservice~Deployment~Problem}
Due to the uncertainty of inference tasks, the deterministic model is hard to meet the situation in the real world because task amount matrix $\mathbf Z^{t}$ is assumed to be known in deterministic modeling. It is necessary to use approaches such as uncertainty modeling and robust optimization to model this dynamic problem. Therefore, we introduce the uncertainty model and the uncertainty set in Section \ref{UM}. The robust optimization of the microservice deployment problem is proposed in Section~\ref{RPF}. 
\label{Section III}

\subsection{Uncertainty Model}
\label{UM}
\begin{figure}[t]
\centering
\includegraphics[width=\linewidth]{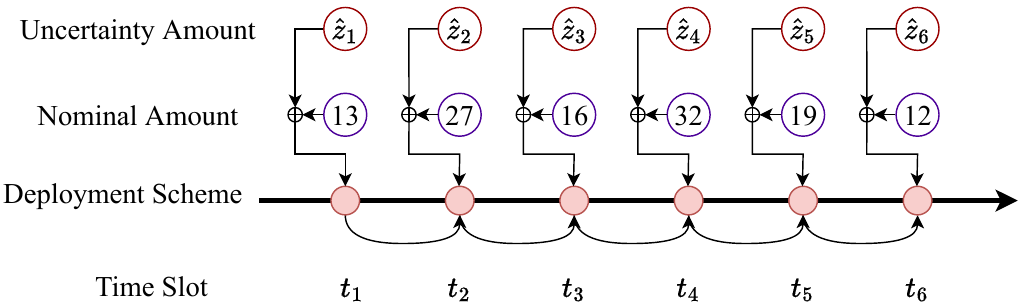}
\caption{Example of data amount uncertainty.}
\label{Uncertainty example}
\vspace{-0.3cm}
\end{figure}

In reality, there can be fluctuations in $z^{t}_{m, s}\in \mathbf{Z}^{t}$ for microservice $m$ and satellite $s$ due to factors such as emergency tasks like earthquake \cite{zhai2015robust}. It can be depicted as Fig. \ref{Uncertainty example}. In particular, the deployment scheme of each time slot is related to 1) the uncertainty data amount, 2) the nominal data amount, and 3) the deployment scheme of the previous time slot. At time slot $t_{2}$, we only know there should be 27 requests in the usual time but have no idea if there are any request fluctuations or not. It may cause significant resource waste or QoS punishment and interfere with the deployment scheme of the following time slots \cite{nguyen2021two}. We introduce uncertainty into the user's request to meet the perturbation of the number of requests. In particular, we build the box uncertainty set on $\mathbf{Z}^{t}$ by
\begin{equation}
\label{uset}  
\Omega(\mathbf{Z}^{t})=\left\{
\mathbf{Z}^{t} \mid \Vert \mathbf{Z}^{t} - \overline{\mathbf Z}^{t} \Vert_\infty \le \Phi
\right\}
\end{equation}
for all $t\in \mathbb{T}$, where $\Phi$ denotes the robustness level (i.e. width of the uncertainty set) of the current model, $\overline{\mathbf{Z}}^{t}$ denotes the nominal request count matrix. Then, we can reformulate the constraint \eqref{yk} into an uncertainty constraint:
\begin{equation}
\label{uyk}
    \min_{\mathbf{Z}^{t}\in \Omega(\mathbf{Z}^{t})}\;\;\sum_{s\in \mathbb{D}} \left\{y_{m,s}^{t}k_{m} - z_{m,s}^{t}\right\}\ge 0
\end{equation}
for all $m \in \mathbb{M}_{l}$, $t\in \mathbb T$.
Constraint \eqref{uyk} indicates that the accessing ability of microservices must be larger than the number of user requests. Due to the difference in each status, simply adding or removing microservices from the deterministic solution won’t resolve this issue due to the different deployment schemes in each time slot. For instance, if $$\sum_{s\in \mathbb{D}}y_{m,s}k_{m}\ge \sum_{s\in \mathbb{D}} (\overline{z}_{m,s}+|\Phi|)$$ for all $\overline{z}_{m,s}\in \overline{\mathbf{Z}^{t}}$, the worst-case scenario is that there aren't enough requests for the microservices to handle, leading to resource wastage and associated penalties. Conversely, if $$\sum_{s\in \mathbb{D}}y_{m,s}k_{m}\textless \sum_{s\in \mathbb{D}} (\overline{z}_{m,s}+|\Phi|),$$ the worst-case scenario shifts to having too many requests to handle, resulting in latency penalties.

\subsection{Robust Microservice Deployment Problem Formulation}
\label{RPF}
Given the QoS constraints \eqref{QoSC} and the robust constraint \eqref{uyk}, we can formulate the robust optimization problem as follows:
\begin{equation}
 \label{obj}   
 \min_{\mathbf{X}\in \mathbb{F}_{1}}\; \mathcal{C}_{1}(\mathbf{X})+\max_{\mathbf{Z}\in\Omega}\min_{\mathbf{Y}\in \mathbb{F}_{2}(\mathbf{X},\mathbf{Z})}\;\mathcal{C}_{2}(\mathbf{Y})+\mathcal{C}_{3}(\mathbf{Y})+\mathcal{C}_{4}(\mathbf{Y}) 
\end{equation}
with $\Omega=\bigcup_{t\in\mathbb T}\Omega(\mathbf Z^t)$.
Here $\mathbf{Z}\in \mathbb{R}^{T\times M_{l}\times D}$ has the request count of all time slot, $\mathbf{Y}\in \mathbb{N}^{T\times M_{l}\times D}$ denotes the light microservice's deployment scheme of all time slot, $\mathbb{F}_{1}$ and $\mathbb{F}_{2}(\mathbf{X},\mathbf{Z})$ is the feasible set of $\mathbf{X}$ and $\mathbf{Y}$ which are shown as:
\begin{align}
    \mathbb{F}_{1}&=\left\{\mathbf{X}\mid \eqref{x1},\eqref{CMRCal} \right\},\\
    \mathbb{F}_{2}(\mathbf{X},\mathbf{Z})&=\left\{\mathbf{Y}\mid \eqref{y1},\eqref{Rstar},\eqref{QoSC},\eqref{EMRCal},\eqref{uyk}\right\}.
\end{align}


While Problem~\eqref{obj} incorporates lots of elements of the real world, such as chain deployment schemes and data amount uncertainty, it raises critical challenges compared with the deterministic works. To show that the optimization problem~\eqref{obj} we proposed is challenging to solve with the simple approach, we will analyze this optimization problem in two parts: NP-hard problem and semi-infinite constraint.
\begin{itemize}
\item \textit{NP-hard Problem}: Suppose that the $\min\max\min$ structure is used in the objective function and $\mathop{\arg\min}$ used in the constraint \eqref{QoSC}, we have the non-convex optimization problem. Moreover, because the total latency function includes a quadratic component, the quadratic constraint (\ref{QoSC}) increases the complexity of solving this optimization problem. Thus, the objective makes it unsolvable in polynomial time by using traditional optimization algorithms.

\item \textit{Semi-Infinite Constraint}: This challenge poses significant interference for global optimization algorithms to solve this problem by introducing an uncertain amount of constraints \eqref{QoSC} into the optimization problem \cite{shi2015robust}. Due to the uncertain set deployed on the number of tasks, the number of quadratic QoS constraints remains unknown. So using the B\&B method to solve this problem is extremely difficult.
\end{itemize}

The analysis reveals that problem \eqref{obj} cannot be efficiently solved by existing optimization algorithms. Sub-optimal methods like greedy algorithms (e.g., K8S's Horizon Pod Autoscaling \cite{s20164621}) are viable but often yield overly conservative strategies (Section \ref{UM}). RL’s Markov property makes it promising for precise decisions without global information. However, deterministic RL struggles with environmental uncertainty. To address this, we propose a robust RL-based approach that retains the Markov property while enhancing robustness.

\section{Robust Adversarial Reinforcement Learning Algorithm}
\label{Section IV}
To address the non-convex robust optimization problem (\ref{obj}) in Section \ref{Section III}, we introduce a robust adversarial reinforcement learning algorithm. Our approach first decomposes the problem (\ref{obj}) into two components based on its inherent properties. We then apply distinct algorithms to solve each part separately.
\subsection{Problem Decomposition}
The traditional two-stage robust optimization framework faces significant challenges due to the high computing complexity caused by the optimal approach such as Benders Decomposition \cite{rahmaniani2017benders}. Due to core microservices' high deployment cost and importance in the problem \eqref{obj}, only $x^{0}_{m,s}$ influence $y^{t}_{m,s}$. Therefore, we can decompose the problem in equation \eqref{obj} into two stages: deploying $\mathbb{M}_{c}$ (\textbf{first stage}) and deploying $\mathbb{M}_{l}$ (\textbf{second stage}). Given the substantial resource consumption of core microservices, each microservice in $\mathbb{M}_{c}$ is deployed only once to avoid excessive resource use. Thus, with the infinite accessing ability, like the computing center, the first stage problem can be transformed into a new form: finding a deployment scheme that minimizes total latency. There are multiple works about how to deploy core microservices in Space-CPN. The first stage's deployment algorithm we adopt is shown as follows.
\subsection{First Stage Deployment Approach}
To address the static microservice deployment problem in the first stage, we adopt an RL-based algorithm to solve it. The standard PPO-based training algorithm is shown in Algorithm~\ref{a1}.
\subsubsection{State and Action}
The state at step $i$ is a vector $\mathbf{s}^{1}_{i}$ where cardinality is $M_{c}\times D$, which means each microservice deploys on each satellite or not. For the first stage, the action $a^{1}_{i}$ at step $i$ is to deploy microservice $i$ on a satellite. With $D$ satellites for each time slot, there are $D$ actions: $a^{1}_{i}\in\{0,1,2,...,D\}$.
\subsubsection{Reward}
If this step or the next step violates the resource limitation, the agent will receive a negative punishment. If the agent successfully deploys one microservice, it will receive a positive reward. When the agent successfully deploys all microservices, it will receive the final reward, which is negatively correlated with the total latency.
\label{FSDA}
\begin{algorithm}[t]
\renewcommand{\algorithmicrequire}{\textbf{Input:}}
\renewcommand{\algorithmicensure}{\textbf{Output:}}
\caption{First Stage Training Process}
\begin{algorithmic}[1]
\REQUIRE{$G_{D} = (\mathbb{D},\mathbb{E})$, $\mathbb{M}_{c} = \{m_{1},m_{2},...,m_{M_{c}}\}$, stochastic policy $\pi_{\theta_{1}}$.}
\ENSURE{Learnable policy parameters $\theta_{1}$.}
\FOR {$episode$ = 1 to $Episode$}
    \STATE $i$ = 1.
    \WHILE {Deployment process not complete} \label{line3} 
    \STATE Get state $\mathbf{s}^{1}_{i}$. \label{line4} 
    \STATE Choose an action $a^{1}_{i}$ with policy $\pi(a^{1}_{i}|\mathbf{s}^{1}_{i},\theta_{1})$.\label{line5} 
    \STATE Execute $a^{1}_{i}$, obtain reward $r_{i}$, generate state $\mathbf{s}_{i+1}$. \label{line6} 
    \STATE $i=i+1$. \label{line7} 
    \ENDWHILE \label{line8} 
    \STATE Rollout $\mathbb{\zeta}$ = \{$(a^{1}_{1},r^{1}_{1}), (a^{1}_{2},r^{1}_{2}),...,(a^{1}_{i-1},r^{1}_{i-1})$\}. \label{line9} 
    \STATE Update policy parameters $\theta_{1}$ with $\mathbb{\zeta}$. \label{line10} 
\ENDFOR \label{line11} 
\end{algorithmic}
\label{a1}
\end{algorithm}

\subsection{Second Stage Deployment Approach}
\subsubsection{Robust Adversarial Reinforcement Learning}
\begin{figure}[t]
\centering
\includegraphics[width=\linewidth]{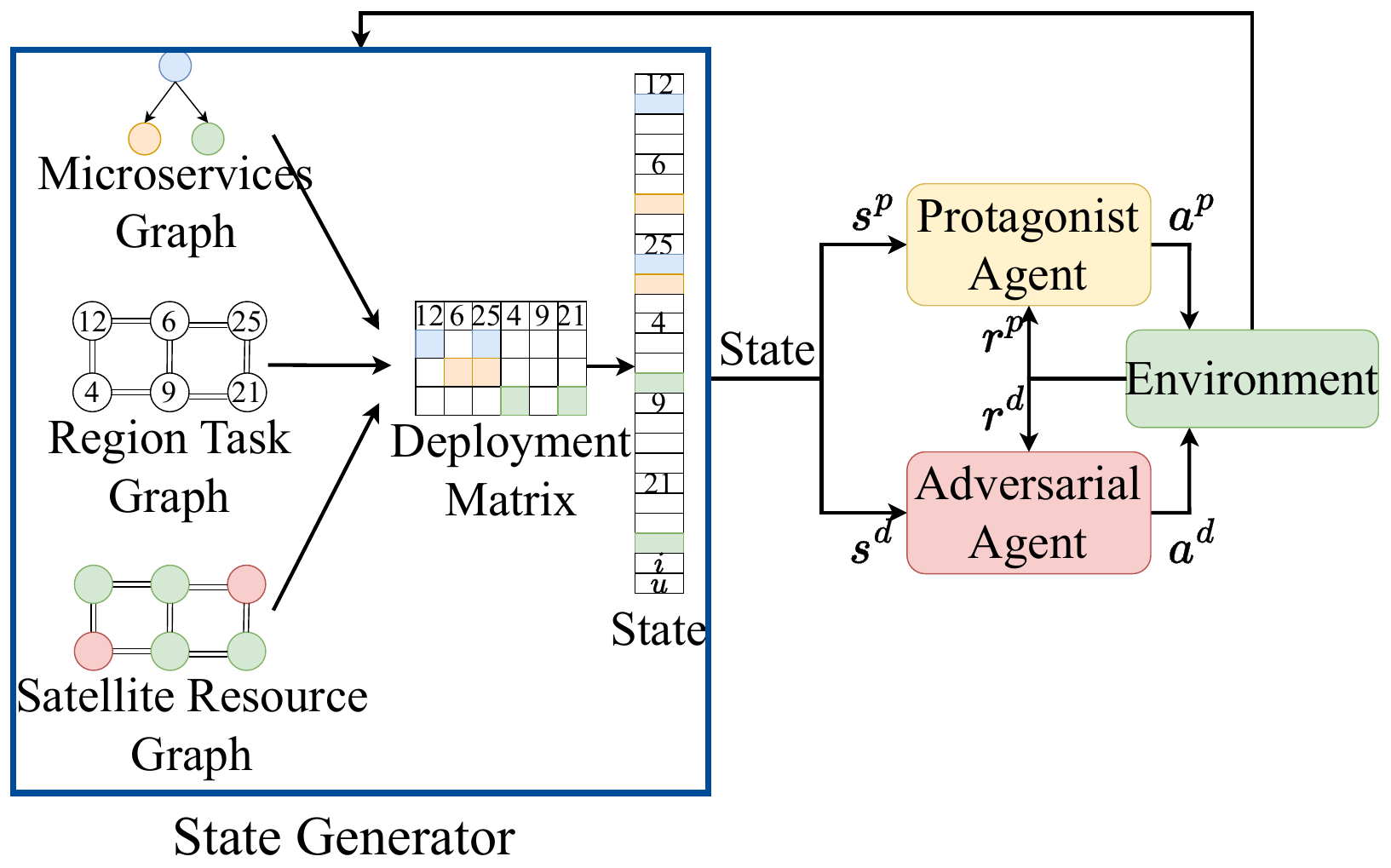}
\caption{Multi-Step Robust Adversarial Reinforcement Learning framework.}
\label{RARL}
\vspace{-0.3cm}
\end{figure}

As discussed in Section \ref{Section III}, simply adjusting the corrections of the uncertainty variable $\mathbf{Z}$ offers limited improvements, as increases and decreases in task volumes cause losses and worst-case scenarios vary by deployment scheme. To address this, we propose a Multi-Step Robust Adversarial Reinforcement Learning (MSRARL) framework using the Proximal Policy Optimization (PPO) algorithm and the Robust Adversarial Reinforcement Learning (RARL) \cite{10.5555/3305890.3305972}, shown in Fig. \ref{RARL}. In this framework, a protagonist agent and an adversary agent interact iteratively. The adversary determines additional requests for each region, while the protagonist makes initial decisions and trains based on current conditions. The adversary then trains using the protagonist's rollout, and the protagonist retrains using the adversary's rollout. This iterative process continues until convergence. Unlike \cite{10.5555/3305890.3305972}, agents can perform multiple actions until their allotted step.

\subsubsection{Partially-Observable Markov Decision Process}
\label{POMDP}
Different from the MDP used in the first stage, the second stage uses the Partially Observable Markov Decision Process (POMDP) due to the perturb request from the environment. POMDP in this paper can be written as a tuple $(\mathbb{S'},\mathbb{S},\mathbb{A}_{1},\mathbb{A}_{2},\mathcal{T},\mathcal{R},\mathcal{O})$ where $\mathbb{S'}$ and $\mathbb{S}$ denote the set of real state and the observation state respectively, $\mathbb{A}_{1}$ and $\mathbb{A}_{2}$ denote the set of action space of the protagonist agent and the adversary agent, $\mathcal{T}:\mathbb{S'}\times \mathbb{A}\rightarrow \mathbb{S'}$, $\mathcal{R}:\mathbb{S'}\times \mathbb{A}_{1}\times\mathbb{A}_{2}\rightarrow \mathbb{R}$, and $\mathcal{O}:\mathbb{S'}\times \mathbb{A}_{1}\times\mathbb{A}_{2}\rightarrow \mathbb{S}$ denote the transition function, reward function, and the observation function. We denote the microservice deployment state and the requests state as matrix $\mathbf P\in\mathbb N^{(1+M_{l})\times D}$ where $(1+M_{l})$ denotes request count and $M_{l}$ microservices, $p_{u,d}$ for all $ u\in \mathbb{M}_{l},d\in \mathbb{D}$ denotes the deployment count of microservice $u$ on satellite $d$ and $p_{0,d}$ is the request count of each region. Then, we stretches the $P$ row by row into a vector as state $\mathbf{s}^{p}_{i}\in \mathbb{N}^{(1+M_{l})D+2}$ at step $i$ given by
\begin{align}
\mathbf{s}^{p}_{i}=[\text{vec}(\mathbf P),i,u],
\end{align}
where $i$ and $u$ denote the protagonist agent's time slot and the current microservice, respectively. The adversary agent's state $\mathbf{s}^{d}_{j}\in \mathbb{N}^{(1+M_{l})D+1}$ also uses marking matrix $P$ to denote the environment, and it is shown as follows:
\begin{align}
\mathbf{s}^{d}_{j}=[\text{vec}(\mathbf{P}_{\text{ori}}),j],
\end{align}
where $j$ denotes the adversary agent's step, $\mathbf{P}_{ori}\in\mathbb N^{(1+M_{l})\times D}$ denotes the $\mathbf{P}$ of previous training round. After the protagonist agent finishes $M_{l}$ steps, the adversary agent will take its next step. To enhance the robustness, we assume the adversary agent can't know the microservice's deployment scheme at step $j$ before it makes a decision.
\subsubsection{Action}
The action space $\mathbb{A}_{1}$ and $\mathbb{A}_{2}$ of the second stage is a discrete space. For the second stage, the action $\mathbf{a}^{p}_{i}$ at step $i$ for microservice $u$ is a vector whose cardinality is $D$. Each element in $\mathbf{a}^{p}_{i}$ is a non-negative integer up to $\alpha$ denotes how many microservice $u$ will be deployed on each satellite where $\alpha$ denotes how many microservices can be deployed on one satellite for a time slot. It was shown as follows:
\begin{align}
\mathbf{a}^{p}_{i}[e]\in \{0,1,2,\dots,\alpha\},e=0,1,2,...,D,
\end{align}
where $[\cdot]$ denotes accessing the vector's element through the index. The action $\mathbf{a}^{d}_{j}$ at step $j$ is a vector whose cardinality is $D$. Each element in $\mathbf{a}^{d}_{j}$ is an integer that denotes the unexpected request from each region, which is bounded by the width of uncertainty set $\Omega(\mathbf{Z^{t}})$: $\Phi$. It was shown as follows:
\begin{align}
\mathbf{a}^{d}_{j}[e]\in \{q\mid q\in \mathbb{Z},|q|\le \Phi\},e=0,1,2,...,D.
\end{align}
\subsubsection{Reward}
We design the protagonist reward function with three parts: microservice deployment reward $r^{m}_{i}$, time slot reward $r^{s}_{i}$, and all time slots reward $r^{a}_{i}$. For each step $i$, the protagonist agent will try to deploy a microservice, and the deployment reward or punishment is shown in the following function:
\begin{align}
\label{sr1}
    r^{m}_{i}=&\left\{
        \begin{array}{ll}
            \epsilon_{1},&\text{deployment violated the resource limitation,} \\
            \epsilon_{2},&\text{successfully deployed one microservice,}
        \end{array}
    \right.
\end{align}
where $\epsilon_{1}<0$ and $\epsilon_{2}>0$ denotes the punishment and reward, respectively. Then, after $M_{l}$ steps, the protagonist agent will finish all microservices deployment of a time slot. Now we use the time set with 1 time slot $\mathbb{T}^{i}=\{t=i / M_{l}\}$ to replace the $\mathbb{T}$ used in the $\mathcal{C}_{2}\;\eqref{o2}\sim\mathcal{C}_{4}\;\eqref{o4}$ and receive the cost function $\mathcal{C}^{t}_{2},\mathcal{C}^{t}_{3}$, and $\mathcal{C}^{t}_{4}$ for time slot $t=i / m$. The protagonist agent will receive the time slot reward $r^{s}_{i}$ as the following definition:
\begin{align}
\label{sr2}
    r^{s}_{i}=&\left\{
        \begin{array}{ll}
            \epsilon_{3}\sum_{k=2}^{4}\mathcal{C}^{t}_{k},& i\;\bmod\;(m\;-\;1)\;=\;0, \\
            0,&\text{others,}
        \end{array}
    \right.
\end{align}
where $\epsilon_{3}<0$ denotes the resource punishment factor. Finally, after all $T$ time slots, the protagonist will receive the final reward $r^{a}_{i}$. It is related to the total cost of microservices \eqref{o2}$\sim$\eqref{o4} and the QoS constraint \eqref{QoSC}. In particular, $r^{a}_{i}$ is shown as:
\begin{align}
\label{sr3}
    r^{a}_{i}=&\left\{
        \begin{array}{ll}
            \epsilon_{4}\sum_{i=0}^{T-1}\sum_{k=2}^{4}\mathcal{C}^{t}_{k}+\epsilon_{5}Q,&i\;=\;mT\;-\;1, \\
            0,&\text{others,}
        \end{array}
    \right.
\end{align}
where $\epsilon_{4}<0$ and $\epsilon_{5}<0$ denotes the resource and QoS punishment factor respectively, $Q$ is the times of QoS constraint \eqref{QoSC} violation. With the sub-reward function \eqref{sr1}, \eqref{sr2}, and \eqref{sr3}, we can define the protagonist reward for step $i$ as
\begin{align}
    r^{p}_{i}=r^{m}_{i}+r^{s}_{i}+r^{a}_{i}.
\end{align}
Due to the adversarial agent only taking one action in each time slot (i.e., one step corresponds to a time slot), the adversarial reward is the negative sum of the reward received by the protagonist agent in a time slot, which is 
\begin{align}
    r^{d}_{j}=-\sum_{u=0}^{M_{l}-1}r^{p}_{jM_{l}+u}.
\end{align}

\subsubsection{Adversarial Training Approach}
The training process of the MSRARL framework is detailed in Algorithm \ref{a2}. We initialize the neural network parameters for the protagonist and adversary agents (line \ref{line2-0}). A state generator is incorporated to integrate microservice deployment details, region task amounts, and remaining resources. These inputs are converted into a deployment matrix and vectorized into a state vector for use by both agents. First, the adversary agent generates perturbation tasks and introduces them into the protagonist agent's environment (lines \ref{line2-3}-\ref{line2-4}). The protagonist agent is then trained in this perturbed environment using a PPO-based policy gradient optimizer (lines \ref{line2-5}-\ref{line2-8}). The rollout operation involves using agents to generate trajectories of states, actions, and rewards, which are then used to update the policy parameters.  After training for $e$ epochs, the protagonist agent is well-trained and can generate deployment schemes (protagonist rollouts). Next, the protagonist agent infers and collects its deployment scheme $\zeta^{p}$, which is used to train the adversary agent (lines \ref{line2-9}-\ref{line2-16}). This iterative process continues until both agents converge.

\begin{algorithm}[t]
\renewcommand{\algorithmicrequire}{\textbf{Input:}}
\renewcommand{\algorithmicensure}{\textbf{Output:}}
\caption{MSRARL Training Process}
\begin{algorithmic}[1]
\REQUIRE{$G_{D} = (\mathbb{D},\mathbb{E}_{2})$, $\mathbb{M}_{l} = \{m_{1},m_{2},...,m_{M_{l}}\}$, stochastic policies $\pi_{\theta^{p}}$ and $\pi_{\theta^{d}}$ for protagonist agent and adversarial agent, epoch size $e$.}
\ENSURE{Learnable policy parameters $\theta^{p},\theta^{d}$.}
\STATE Init the learnable parameters of protagonist agent $\theta^{p}$ and adversary agent $\theta^{d}$. \label{line2-0}
\FOR {$i$ = 1 to $Episode$} \label{line2-1} 
    \STATE $\theta^{p}_{i}\leftarrow \theta^{p}_{i-1}.$ \label{line2-2} 
    \STATE Collecting adversary rollout $\zeta^{d}$ by using $\theta_{d}$. \label{line2-3} 
    \STATE Adding $\zeta^{d}$ into protagonist agent's environment. \label{line2-4} 
    \FOR{$i$=1,2,\dots,$e$} \label{line2-5} 
        \STATE $\{(\mathbf{s}^{p},\mathbf{s}^{d},\mathbf{a}^{p},\mathbf{a}^{d},r^{p},r^{d})\}\leftarrow \text{rollout}(\zeta^{d},\theta^{p}_{i},\theta^{d}_{i-1})$ \label{line2-6}. 
        \STATE $\theta^{p}_{i}\leftarrow \text{PPO-PolicyOptimizer}(\mathbf{s}^{p},\mathbf{a}^{p},r^{p})$. \label{line2-7} 
    \ENDFOR \label{line2-8} 

    \STATE $\theta^{d}_{i}\leftarrow \theta^{d}_{i-1}.$ \label{line2-9} 
    \STATE Collecting adversary rollout $\zeta^{p}$ by using $\theta^{p}_{i}$. \label{line2-10} 
    \STATE Adding $\zeta^{p}$ into protagonist agent's environment. \label{line2-11} 
    \FOR{$i$=1,2,\dots,$e$} \label{line2-12} 
        \STATE $\{(\mathbf{s}^{p},\mathbf{s}^{d},\mathbf{a}^{p},\mathbf{a}^{d},r^{p},r^{d})\}\leftarrow \text{rollout}(\zeta^{p},\theta^{p}_{i-1},\theta^{d}_{i}).$ \label{line2-13} 
        \STATE $\theta^{d}_{i}\leftarrow \text{policyOptimizer}(\mathbf{s}^{d},\mathbf{a}^{d},r^{d})$. \label{line2-14} 
    \ENDFOR \label{line2-15} 
\ENDFOR \label{line2-16} 
\end{algorithmic}
\label{a2}
\end{algorithm}
\vspace{-0.5cm}
\subsection{Equilibrium Analysis}
\label{EA}
This section analyzes the optimal strategy for the proposed game, which is not a standard two-player zero-sum Markov game. 
\begin{lemma}
\label{tpzs}
The adversarial game in Section \ref{POMDP} can be transferred into a two-player zero-sum game.
\end{lemma}
\textbf{Proof.} The protagonist agent can make decisions without the adversary agent because the adversary agent and the protagonist agent have no information about each other except the reward. Moreover, there is no action from the adversary agent during the $M$ protagonist round in a time slot. Therefore, we can compress the $M$ protagonist action into one protagonist action $\mathbf{Y}^{t}\in \mathbb{N}^{M\times D}$ corresponding to all microservices' deployment schemes in time slot $t$. 
\hfill $\blacksquare$
\smallskip

Lemma \ref{tpzs} shows how to transfer our proposed framework into a standard two-player zero-sum game. To prove that a minimax equilibrium exists, we need the following assumption.
\begin{assumption}
\label{ass1}
    Player 1 has $m$ strategies in one round, and player 2 has $n$ strategies. Player 1 respective to protagonist action set $\mathbb{Y}^{t}$, which starts the action first.
\end{assumption}

Assumption \ref{ass1} gives the width and the order of the proposed game. To prove the existence of Nash equilibrium shown in Lemma \ref{mmem}, we must first know the relationship between $v_{1}$ and $v_{2}$ which is shown in Lemma \ref{v1v2}. To prove the Lemma \ref{v1v2}, we have to prove the Lemma \ref{lele}, which shows the saddle point of the game first.
\label{lone}

\begin{lemma}
\label{lele}
    If $v_{1}=v_{2}$, then $\exists \mathbf{y}^{*}\in \mathbb{Y}^{t},\mathbf{z}^{*}\in \mathbb{Z}^{t}$ such that $\forall \mathbf{y}\in \mathbb{Y}^{t},\mathbf{z}\in \mathbb{Z}^{t}$, with:
    \begin{align}
        \sum_{i=1}^{m}\sum_{j=1}^{n}c_{ij}y_{i}z^{*}_{j}\le \sum_{i=1}^{m}\sum_{j=1}^{n}c_{ij}y^{*}_{i}z^{*}_{j}\le \sum_{i=1}^{m}\sum_{j=1}^{n}c_{ij}y^{*}_{i}z_{j}.
    \end{align}
\end{lemma}
\textbf{Proof.} Now we assume $v=v_{1}=v_{2}$. Then $\exists \mathbf{y}^{*}\in \mathbb{Y}^{t},\mathbf{z}^{*}\in \mathbb{Z}^{t}$, we have
\begin{align}
    \max_{\mathbf{y}\in \mathbb{Y}^{t}}\sum_{i=1}^{m}\sum_{j=1}^{m}c_{ij}y_{i}z_{j}^{*}=v=\min_{\mathbf{z}\in \mathbb{Z}^{t}}\sum_{i=1}^{m}\sum_{j=1}^{n}c_{ij}y_{i}^{*}z_{j}.
\end{align}
Therefore, $\forall \mathbf{y}\in \mathbb{Y}^{t},\mathbf{z}\in \mathbb{Z}^{t}$, we have
\begin{align}
    \sum_{i=1}^{m}\sum_{j=1}^{n}c_{ij}y_{i}z_{j}^{*}\le v \le \sum_{i=1}^{m}\sum_{j=1}^{n}c_{ij}y_{i}^{*}z_{j}.
\end{align}
\hfill $\blacksquare$

With the saddle point introduced in Lemma \ref{lele}, we can prove $v_{1}\le v_{2}$ which is shown in Lemma \ref{v1v2}.

\begin{lemma}
\label{v1v2}
    For given reward matrix $\mathbf{C}$, time slot $t$'s strategy set $\mathbb{Y}^{t}$ and $\mathbb{Z}^{t}$, and $v_{1}$ and $v_{2}$ we introduced in Section \ref{PAN}, $v_{1}\le v_{2}$.
\end{lemma}
\textbf{Proof.} We assume player 1 has $m$ strategies and player 2 has $n$ strategies. $\forall \mathbf{y}\in \mathbb{Y}^{t},\mathbf{z}\in \mathbb{Z}^{t}$. According to lemma \ref{lele}, we have
\begin{align}
    \min_{\mathbf{z}\in \mathbb{Z}^{t}}\sum_{i=1}^{m}\sum_{j=1}^{n}c_{ij}y_{i}z_{j}\le
    \sum_{i=1}^{m}\sum_{j=1}^{n}c_{ij}y_{i}^{*}z_{j}^{*}\le
    \max_{\mathbf{y}\in \mathbb{Y}^{t}}\sum_{i=1}^{m}\sum_{j=1}^{n}c_{ij}y_{i}z_{j}.
\end{align}
Therefore, we can say for $\forall t\in \mathbb{T}$, $v_{1}\le v_{2}$.
\hfill $\blacksquare$
\smallskip

With $v_{1}\le v_{2}$ shown in Lemma \ref{v1v2}, to prove the existence of Nash equilibrium shown in Lemma \ref{mmem}, we need to show it is impossible to let $v_{1}\;\textless\; v_{2}$ for the Nash equilibrium point, which is shown in Lemma \ref{2ineq}. To prove this, we must first prove Lemma \ref{wx}.

\begin{lemma}
    \label{wx}
    Suppose $\mathbb{H}\subset \mathbb{R}^{n}$ is a convex and compact set and $\mathbf{0}\notin \mathbb{H}$, then there $\exists \mathbf{w}\in \mathbb{R}^{n}$ such that $\forall \mathbf{x}\in \mathbb{H},\mathbf{w}\cdot \mathbf{x}\textgreater 0.$
\end{lemma}
\textbf{Proof.} There exists a $\mathbf{w}\in \mathbb{H}$ satisfies $\forall \mathbf{x}\in \mathbb{H}, \lvert \mathbf{w} \rvert \le \lvert \mathbf{x} \rvert$. Due to $\mathbb{H}$ is a convex set, $\forall \alpha\in (0,1),\mathbf{x}\in \mathbb{H}, (1-\alpha)\mathbf{w}+\alpha\mathbf{x}\in \mathbb{H}$. Then, we have $\lvert \mathbf{w} \rvert \le (1-\alpha)\mathbf{w}+\alpha\mathbf{x}$. We expand this inequality and take the limit at $\alpha \rightarrow 0$, then we have $\lvert \mathbf{w} \rvert^{2} \le \mathbf{w}\cdot\mathbf{x}$. Due to $\lvert \mathbf{w} \rvert \textgreater 0$, we have $\mathbf{w}\cdot \mathbf{x}\textgreater 0.$
\hfill $\blacksquare$
\smallskip

With the Lemma \ref{wx}, we can start to prove the lemma \ref{2ineq}. Lemma \ref{2ineq} provides two inequalities to prove $v_{1}=v_{2}$ which are used in Lemma \ref{mmem}.

\begin{lemma}
\label{2ineq}
    For given matrix $\mathbf{C}$, at least one of the following two inequalities holds true:
    \begin{align}
        \sum_{i=1}^{m}c_{ij}y_{i}\ge 0,\;j=1,2,...,n,\label{ineq1}\\
        \sum_{j=1}^{n}c_{ij}z_{j}\le 0,\;i=1,2,...,m.\label{ineq2}
    \end{align}
\end{lemma}
\textbf{Proof.} Now we consider $m+n$ nodes: $\mathbf{c}(i)=(c_{i1},c_{i2},...,c_{in}), \forall i\in [1,m]$, $\mathbf{e}(j)=(e_{1},e_{2},...,e_{n})$ where $\forall j\in [1,n], e_{j}=\delta_{jk},$
\begin{equation}
\delta_{jk}=\left\{
\begin{aligned}
-1 & , & \text{if}\;k=j, \\
0 & , & \text{otherwise.}
\end{aligned}
\right.
\end{equation}
We use $\mathbb{H}\in \mathbb{R}^{m}$, a non-empty bounded closed set, to denote the convex hull of these $m+n$ nodes. If $\mathbf{0}\in \mathbb{H}$, there exist $m+n$ non-negative scalars $\alpha_{1},\alpha_{2},...,\alpha_{m},\beta_{1},\beta_{2},...,\beta_{n}\in \mathbb{R}^{+}$ satisfy $\sum_{i=1}^{n}\beta_{i}+\sum_{j=1}^{m}\alpha_{j}=1$ and 
\begin{align}
    \sum_{i=1}^{n}\beta_{i}\mathbf{e}(i)_{i}+\sum_{j=1}^{m}\alpha_{j}c_{jk}=0,
\end{align}
where $\mathbf{e}(i)_{i}$ denotes the $i$-th element of $\mathbf{e}(j)$. We know $\sum_{j=1}^{m}\alpha_{j}\textgreater 0$, otherwise $\sum_{i=1}^{n}\beta_{i}=0$, which contradicts $\alpha$ and $\beta$'s sum constraint. Let $\mathbf{y}=(\alpha_{1},...,\alpha_{m}), \sum_{i=1}^{m}c_{ij}y_{i}=\sum_{j=1}^{n}-\mathbf{e}(j)_{j}\ge 0$. If $\mathbf{0}\notin \mathbb{H}$, according to the Lemma \ref{wx}, there $\exists \mathbf{w}\in \mathbb{R}^{n}$ such that $\forall \mathbf{x}\in \mathbb{H}, \mathbf{w}\cdot \mathbf{x}\textgreater 0$. We first assume $\mathbf{x}=\mathbf{e}(i)$, then we have $\mathbf{w}\cdot \mathbf{e}(i)\textgreater 0$. Then we can say each element in $\mathbf{w}$ is negative. Then we assume $\mathbf{x}=\mathbf{c}(j)$, we have $\mathbf{w}\cdot \mathbf{c}(j)\textgreater 0$. Suppose $\mathbf{z}=-\mathbf{w}$, we have $\sum_{j=1}^{n}c_{ij}z_{j}\le 0$.
\hfill $\blacksquare$
\smallskip

With two inequalities shown in Lemma \ref{2ineq}, we can prove the existence of Nash equilibrium in each matrix game with the mixed strategy which is shown in \ref{mmem}.

\begin{lemma}
\label{mmem}
    Any matrix game has the minimax equilibrium in the sense of mixed strategy.
\end{lemma}
\textbf{Proof.} Due to the policy sampling operation used by PPO, each agent in this game has a strategy probability distribution. In other words, this game is a matrix game. According to Lemma~\ref{v1v2}, we have $v_{1}\le v_{2}$ where $v_{1}$ and $v_{2}$ is shown in Section \ref{PAN}. To show $v_{1}=v_{2}$, we assume $v_{1}\textless a \textless v_{2}$ first. Let $c^{'}_{ij}=c_{ij}-a$, we have $v^{'}_{1}=v_{1}-a\textless 0,v^{'}_{2}=v_{2}-a\textgreater 0$. Based on Lemma~\ref{2ineq} with inequality \eqref{ineq1}, we have 
\begin{align}
    v^{'}_{1}=\max_{\mathbf{y}\in \mathbf{Y}^{t}}\min_{\mathbf{z}\in \mathbf{Z}^{t}}\sum_{i=1}^{m}\sum_{j=1}^{n}c_{ij}y_{i}z_{j}\ge 0,
\end{align}
and with inequality \eqref{ineq2}, we have
\begin{align}
    v^{'}_{2}=\min_{\mathbf{z}\in \mathbf{Z}^{t}}\max_{\mathbf{y}\in \mathbf{Y}^{t}}\sum_{i=1}^{m}\sum_{j=1}^{n}c_{ij}y_{i}z_{j}\le 0.
\end{align}
Both are contradictory to our assumption. Thus, $v_{1}=v_{2}$, i.e., we have the minimax equilibrium in this matrix mixed strategy game.
\hfill $\blacksquare$
\smallskip

Based on Lemma \ref{tpzs} and \ref{mmem}, one can conclude that there exists an optimal strategy point as shown in Theorem \ref{the1}.
\begin{theorem}
\label{the1}
The optimal strategy of minimax equilibrium and Nash equilibrium are equivalent to the two-player zero-sum game transferred from Section \ref{POMDP}.
\end{theorem}
\textbf{Proof.} According to Lemma \ref{tpzs}, our adversarial game can be transferred into a two-player zero-sum game. Moreover, according to Lemma \ref{mmem}, we have the minimax equilibrium in this game. Due to the property of the two-player zero-sum game, the minimax equilibrium is equal to the Nash equilibrium (i.e., the optimal strategy for two players) in the proposed game~\cite{10.5555/3305890.3305972}, \cite{pmlr-v37-perolat15}. 
\hfill $\blacksquare$
\smallskip


\section{Simulation~Results}
\label{Section V}
In this section, we evaluate the performance of the proposed robust microservice deployment algorithm in terms of accessing request count, count of microservices, and computational complexity.
\subsection{Simulation Settings}
In the following experiments. we designed the LEO satellite network, which has limited resources on each satellite edge node, into the mesh topology and randomly generated distance between satellites. For example, a node has 4 CPU cores, 4 GB of memory, 4 GPU cores, and 200 W power \cite{wang2020mpcsm}. The onboard microservice edge AI remote sensing inference system is deployed in these satellite nodes. We used 6 satellites and 2 microservices in our experiment to determine performance gaps between baselines and ours. Nominal requests from each region will be set as an integer from $[0,30]$. To evaluate the performance of our proposed microservice deployment algorithm, we use three deployment algorithms as a baseline to compare the performance and the computation complexity, as shown below.
\begin{itemize}
\item \textbf{\textit{Vanilla RL}}: the policy gradient agent chooses the deployment scheme without uncertainty set.
\item \textbf{\textit{Heuristic Method}}: the deployment scheme is chosen by K8S's Horizon Pod Autoscaling (HPA), i.e., deploy one more microservice if the satellite has too many requests \cite{s20164621}.
\item \textbf{\textit{Robust Heuristic Method}}: the deployment scheme is chosen by K8S's HPA strategy with uncertainty correction.
\end{itemize}
\subsection{Performance of Microservice Deployment Algorithm}
\begin{figure}[t]
\centering
\includegraphics[width=0.88\linewidth]{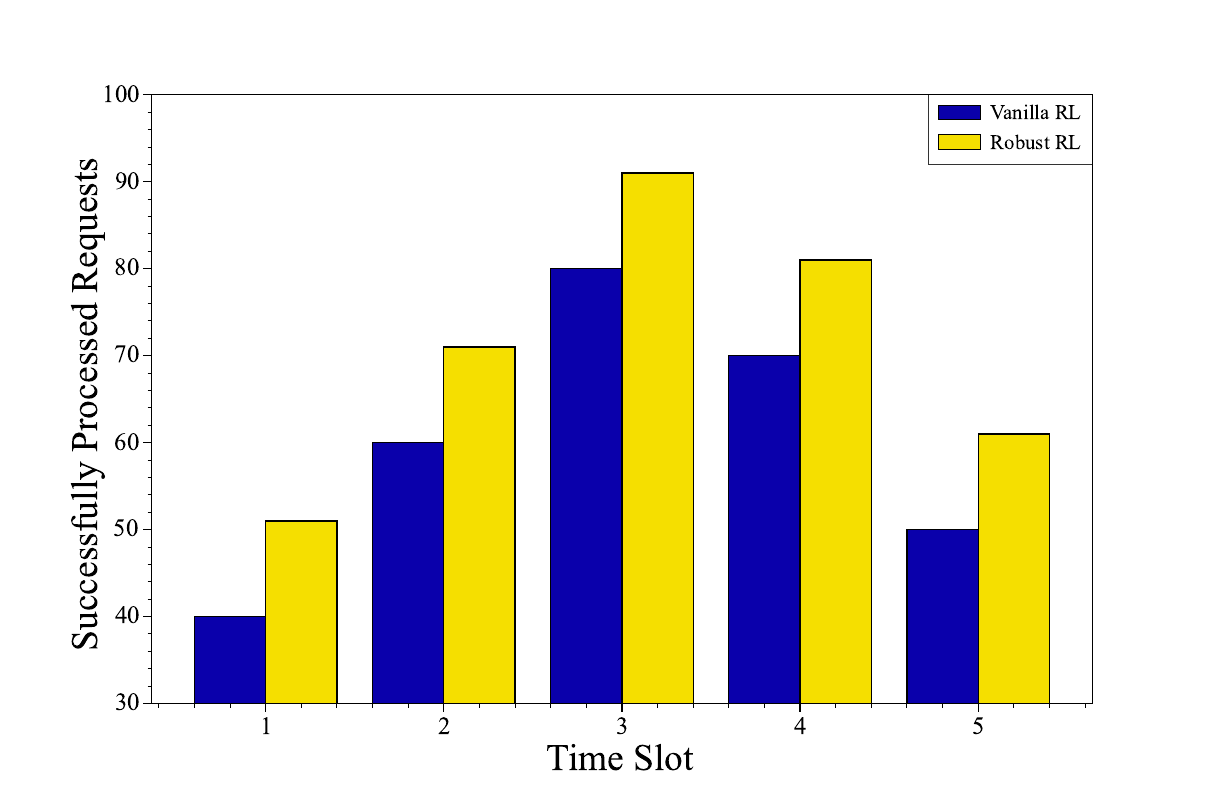}
\caption{Successfully processed requests between vanilla RL and robust RL.}
\label{Successfully Processed Request}
\vspace{-0.3cm}
\end{figure}
\begin{figure}[t]
\centering
\includegraphics[width=0.8\linewidth]{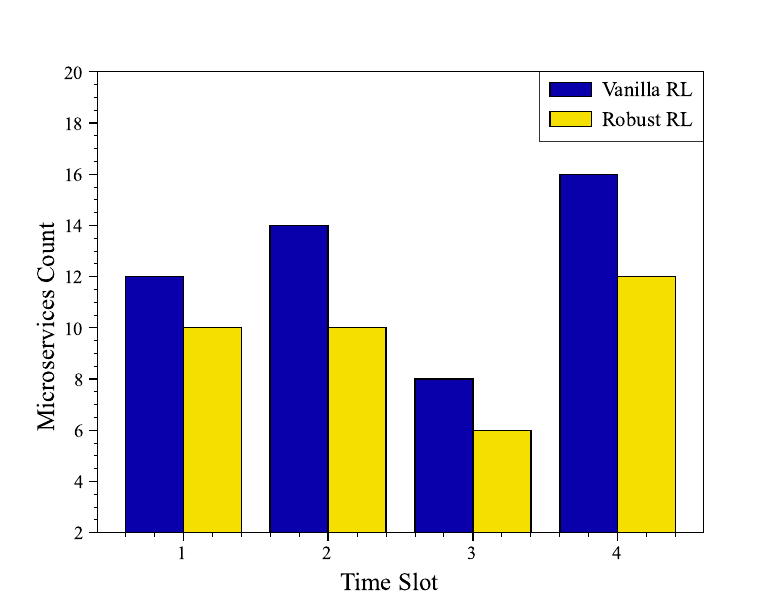}
\caption{Microservice deployment count between vanilla RL and robust RL.}
\label{Microservice Deployment Count 1}
\vspace{-0.3cm}
\end{figure}

According to Section \ref{Section III}, we have different robust optimal solutions for different request states. To show that our proposed algorithm can solve both situations better than the vanilla RL algorithm, we divide all time slots into two parts: (1) need more microservices and (2) need fewer microservices. We first deploy microservices for situation (1) with vanilla RL and robust RL we proposed. We apply the requested amount of each time slot as 55, 65, 27, 87, and 76, respectively. The width uncertainty set used in Fig. \ref{Successfully Processed Request} and Fig. \ref{Microservice Deployment Count 1} is 2. For each time slot in Fig. \ref{Successfully Processed Request}, our proposed algorithm processed more tasks than vanilla RL by deploying more microservices. For another situation, we deploy microservices by using two algorithms, and the result is shown in Fig. \ref{Microservice Deployment Count 1}. We apply the requested amount of each time slot as 43, 54, 63, and 54, respectively. For each time slot, our proposed algorithm can recognize the situation that needs to deploy fewer microservices than usual and reduce the amount of deployment to decrease the resource consumption of microservices. 
\begin{figure}[t]
\centering
\includegraphics[width=0.85\linewidth]{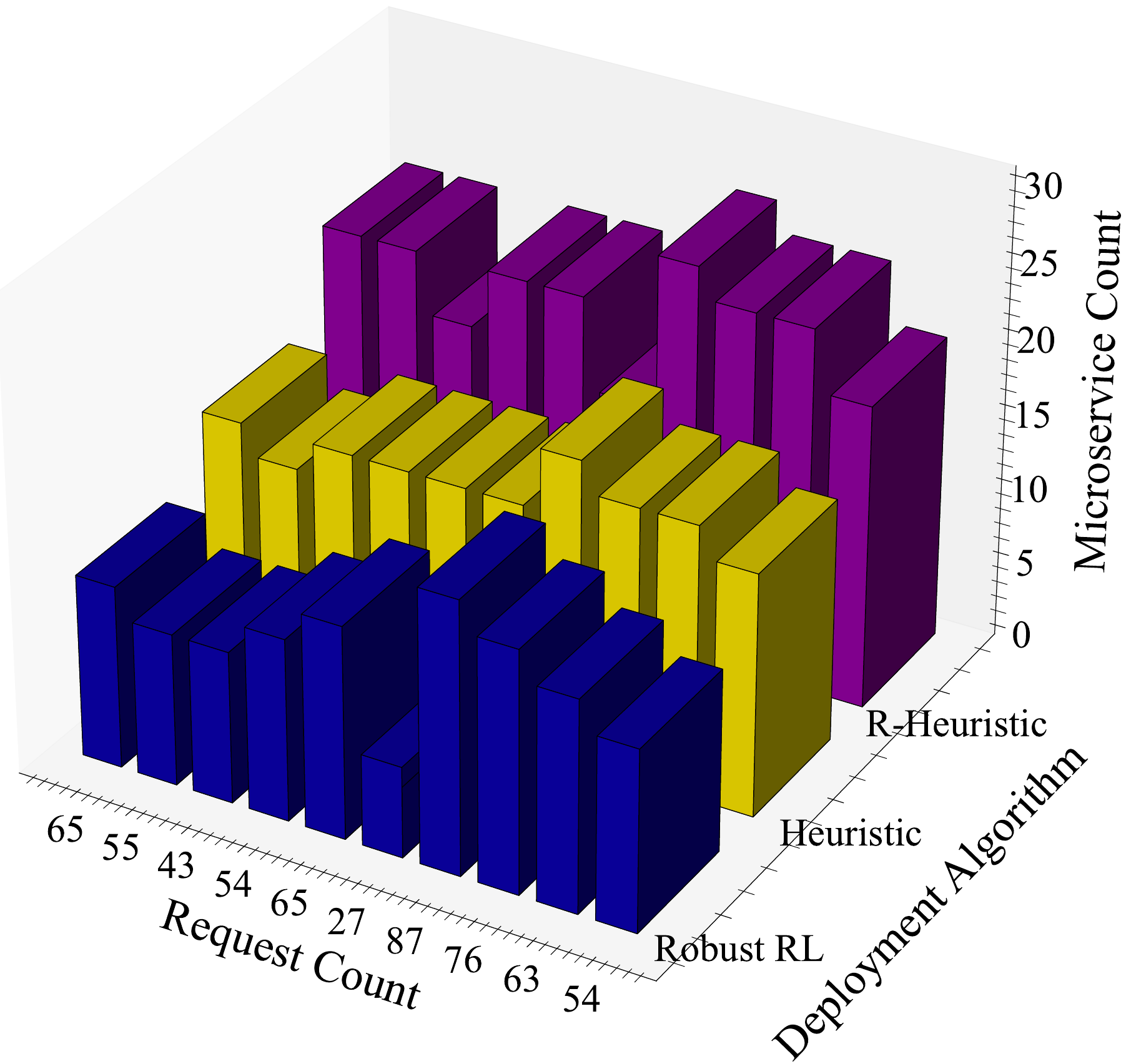}
\caption{Microservice count of MSRARL and Heuristic.}
\label{RARL V.S. Heuristic}
\vspace{-0.3cm}
\end{figure}

Then, we compare the performance of the heuristic algorithm, robust heuristic algorithm, and our proposed algorithm. Due to K8S's default deployment strategy already having limited robustness, we compare the resource consumption (i.e., the amount of microservices deployment) to show our algorithm's advantage and apply the width of uncertainty set as 1. The result in Fig. \ref{RARL V.S. Heuristic} indicates that our proposed algorithm outperforms the robust result obtained by the heuristic algorithm and a robust heuristic algorithm. In other words, for two situations, our proposed algorithm can obtain less deployment than heuristic algorithms. The over-robustness of the heuristic method may be the reason for our proposed algorithm's outperforming. It will deploy more microservices if there are too many requests on each satellite, but each satellite can send overflow requests to other satellites by using ISLs.

\begin{figure}[t]
\centering
\includegraphics[width=\linewidth]{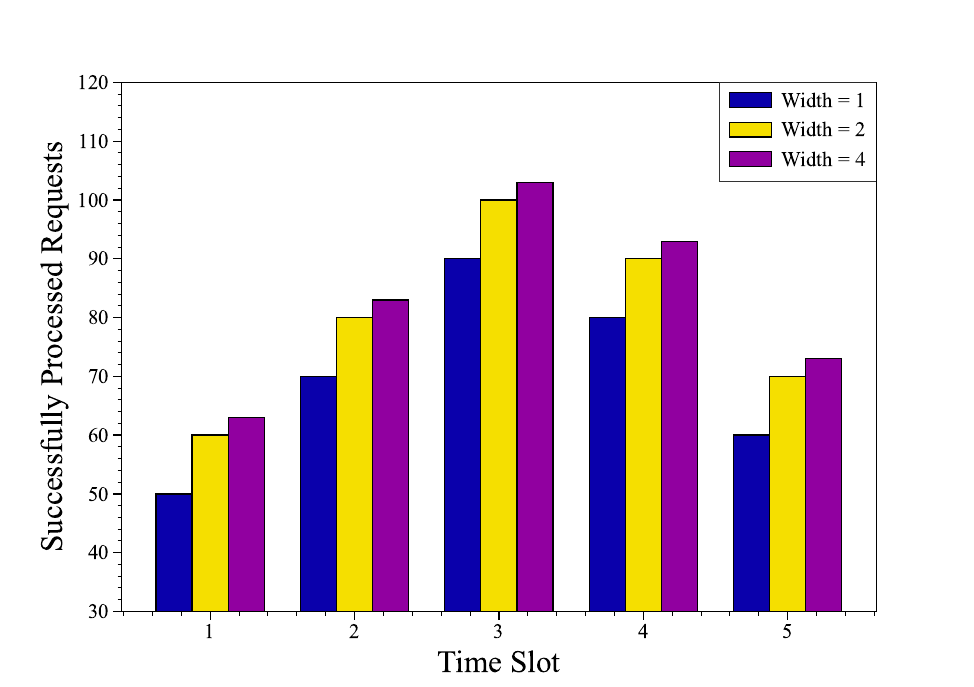}
\caption{Request counts between different uncertainty set widths.}
\label{Successfully Processed Request 2}
\end{figure}

\begin{figure}[t]
\centering
\includegraphics[width=0.95\linewidth]{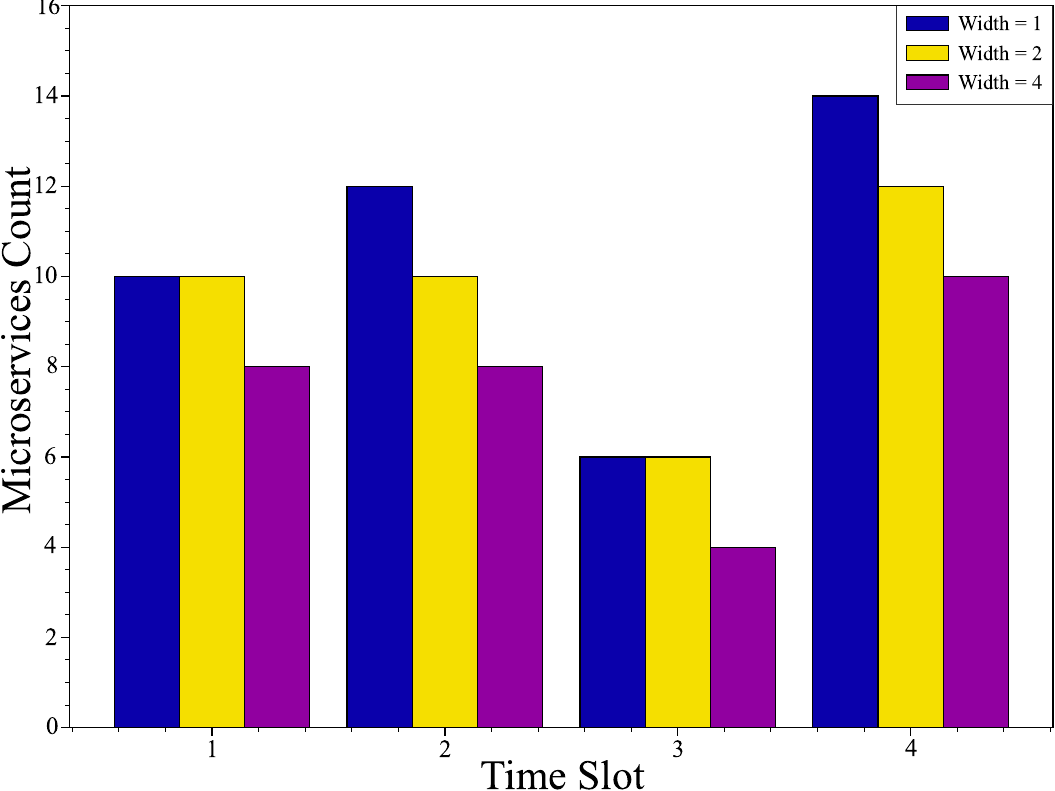}
\caption{Microservice counts between different uncertainty set widths.}
\label{Microservice Deployment Count 2}
\vspace{-0.3cm}
\end{figure}

\begin{figure}[t]
\centering
\includegraphics[width=\linewidth]{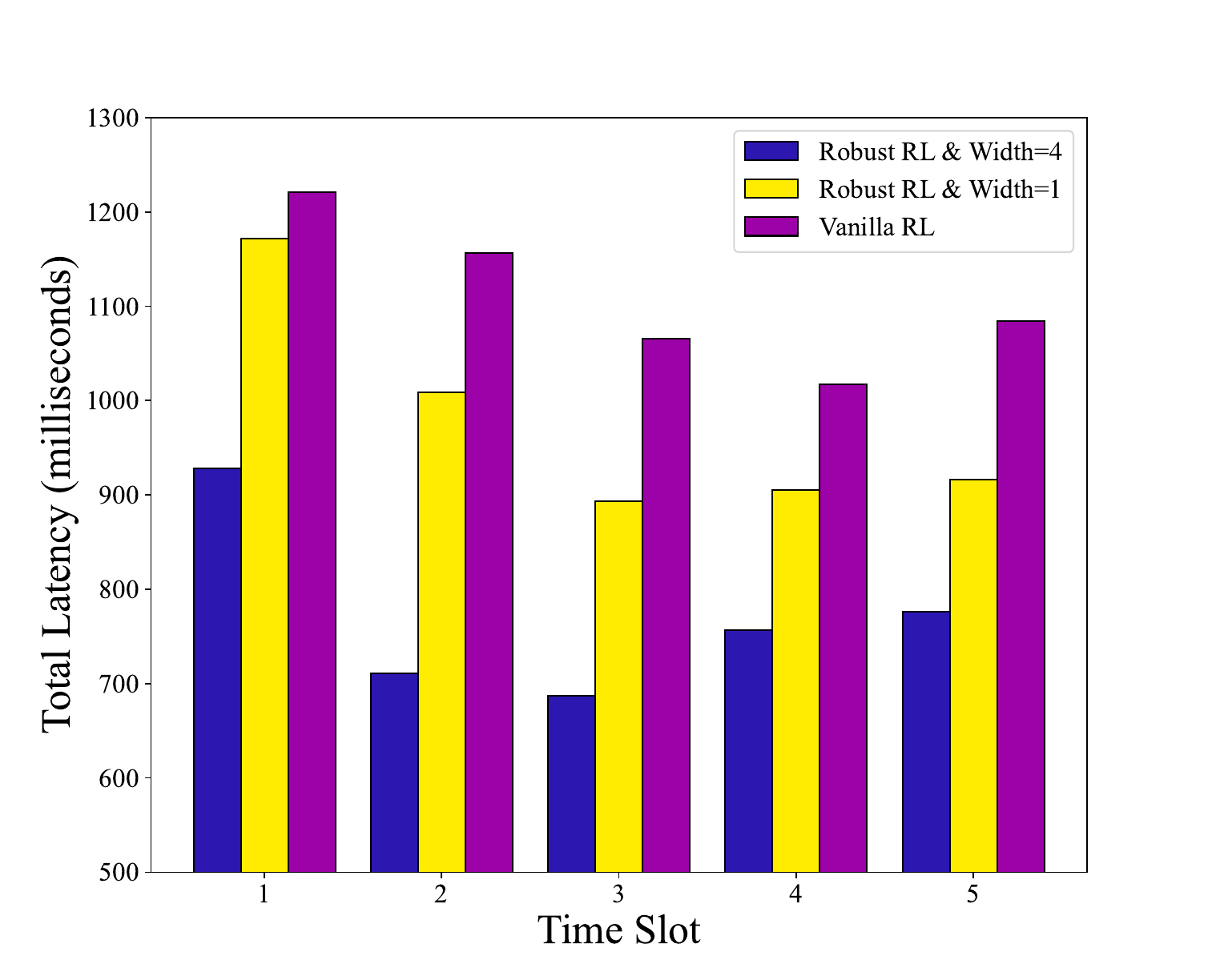}
\caption{Service total latency of robust RL and vanilla RL.}
\label{Service total latency}
\vspace{-0.3cm}
\end{figure}

\begin{table*}[t]
\centering
\renewcommand{\arraystretch}{1.5}
\caption{Running time of Robust RL, Heuristic, and Robust Heuristic under different problem scales.}
\label{table1}
\begin{tabular}{|c|c|c|c|c|c|c|}
\hline
\multirow{2}{*}{\textbf{Algorithms}} & \multicolumn{3}{c|}{\textbf{2 Microservices (millisecond)}}&\multicolumn{3}{c|}{\textbf{4 Microservices (millisecond)}}\\
\cline{2-7}

& \textbf{6 Satellites} & \textbf{12 Satellites} & \textbf{18 Satellites} & \textbf{6 Satellites} & \textbf{12 Satellites} & \textbf{18 Satellites} \\
\hline

Robust RL &\num{9.594e2}	&\num{9.624e2}	&\num{9.897e2}	&\num{9.737e2}	&\num{1.021e3}	&\num{1.075e3}\\ 
\hline

Heuristic & \num{4.673e-2} & \num{8.845e-2} &	\num{1.149e-1} &	\num{4.863e-2} &	\num{8.821e-2} &	\num{1.161e-1}\\
\cline{2-7}
\hline

Robust Heuristic &\num{4.816e-2}	&\num{9.56e-2}	&\num{1.223e-1}	&\num{4.982e-2}	&\num{8.654e-2}	&\num{1.218e-1}\\

\hline
\end{tabular}
\label{t2}
\end{table*}

To show the relationship between the width of the uncertainty set and the deployment scheme, we apply three different widths (1, 2, and 4) of the uncertainty set and the trained policy network, respectively. Then, we use 4 as the width of the uncertainty set for each request region. The result of the two situations is shown in Fig. \ref{Successfully Processed Request 2} and Fig. \ref{Microservice Deployment Count 2}. Fig. \ref{Successfully Processed Request 2} indicates that we can increase the amount of successfully processed requests by deploying more microservices at each time slot. With the increasing width of the uncertainty set, the inference system can process more tasks because the agent uses different uncertainty sets during training. Fig. \ref{Microservice Deployment Count 2} indicates that for each time slot that needs a decreased amount of microservice, we can get a more robust solution by increasing the width of the uncertainty set. On the other hand, if we want to receive a more robust solution by increasing the width of the uncertainty set, we may consume more resources to deploy more microservices. Additionally, to demonstrate the improvement in the overall service latency of our framework, we compare the latency between vanilla RL and robust RL with uncertainty set widths of 1 and 4, as shown in Fig. \ref{Service total latency}. For each time slot, when multiple unexpected tasks need to be performed, our framework achieves lower average latency than the vanilla RL. Furthermore, as the uncertainty set width increases, the total latency decreases. This phenomenon can be attributed to the fact that a larger uncertainty set allows the algorithm to deploy more microservices, enabling it to handle more unexpected tasks efficiently. 

The computation complexity of the robust RL, heuristic method, and robust heuristic method in different scenarios are shown in Table \ref{table1}. The running time is obtained by deploying 2 and 4 microservices into a satellite network with 6, 12, and 18 satellites using these three deployment algorithms. As shown in Table \ref{table1}, the heuristic method and robust heuristic method have shorter computation times than our proposed method because the decision-making process is simpler. However, our proposed algorithm’s computation time for each time slot is less than 110 ms, which can meet the fast on-board deployment requirement.

\section{Conclusion}
\label{Section VI}
This paper proposed a robust reinforcement learning framework for efficient microservice deployment in satellite-based remote sensing. Leveraging a microservice architecture optimizes resources used in LEO satellite networks while addressing heterogeneous resource constraints. The deployment problem, modeled as a two-stage robust optimization, minimized resource consumption while meeting QoS and resource requirements. The problem was decomposed into sub-problems and solved using a robust reinforcement learning algorithm to handle complexity from semi-infinite constraints. Experiments showed that the framework outperforms baselines like standard reinforcement learning, heuristics, and robust heuristics, delivering better performance within acceptable times.

\bibliographystyle{IEEEtran}
\bibliography{reference}

\clearpage

\end{document}